\documentclass[%
preprint,
bibnotes,
 amsmath,amssymb,
 aps,
]{revtex4-2}

\usepackage{graphicx}
\usepackage{dcolumn}
\usepackage{bm}

\usepackage{epstopdf, epsfig}
\usepackage{amsfonts}
\usepackage{mathrsfs}
\usepackage{color}

\newcommand\beq{\begin{equation}}
\newcommand\eeq{\end{equation}}
\newcommand\beqa{\begin{eqnarray}}
\newcommand\eeqa{\end{eqnarray}}
\newcommand{\ud}{\mathrm{d}}
\def\half {\textstyle{\frac{1}{2}}}

\newcommand{\irm}{\mathrm{i}}

\begin{document}


\title{Stabilisation of exact coherent structures in two-dimensional turbulence using time-delayed feedback}

\author{Dan Lucas}
\author{Tatsuya Yasuda}
\affiliation{School of Computing and Mathematics, Keele University, Staffordshire, ST5 5BG, UK}
\email{d.lucas1@keele.ac.uk}

\date{\today}

\begin{abstract}
Time-delayed feedback control, attributed to Pyragas (1992 \emph{Physics Letters \textbf{170}, 421}), is a method known to stabilise periodic orbits in low dimensional chaotic dynamical systems. A system of the form $\dot{\bm{x}}(t)=f(\bm{x})$ has an additional term $G(\bm{x}(t-T)-\bm{x}(t))$ introduced where $G$ is some `gain matrix' and $T$ a time delay. The form of the delay term is such that it will vanish for any orbit of period $T,$ therefore making it also an orbit of the uncontrolled system. This noninvasive feature makes the method attractive for stabilising exact coherent structures in fluid turbulence. Here we begin by validating the method for the basic flow in Kolmogorov flow; a two-dimensional incompressible Navier-Stokes flow with a sinusoidal body force. The linear predictions for stabilisation are well captured by direct numerical simulation. By applying an adaptive method to adjust the streamwise translation of the delay, a known travelling wave solution is able to be stabilised up to relatively high Reynolds number. We discover that the famous `odd-number' limitation of this time-delayed feedback method can be overcome in the fluid problem by using the symmetries of the system. This leads to the discovery of eight additional exact coherent structures which can be stabilised with this approach. This means that certain unstable exact coherent structures can be obtained by simply time-stepping a modified set of equations, thus circumventing the usual convergence algorithms. 
\end{abstract}

\maketitle

\section{Introduction}

 Borrowing mathematical theory from dynamical systems and applying it to the Navier-Stokes equations has seen the computational discovery of unstable exact coherent structures (ECSs) which serve as organising centres of a turbulent flow. These unstable solutions can take the form of steady equilibria, travelling waves or time periodic orbits. The idea is that chaotic trajectories navigate a high dimensional phase space between the neighbourhoods of these solutions directed via their stable and unstable manifolds \citep{Kawahara:2001ft,Gibson:2008ec,vanVeen:2011cm,Kawahara:2012iu}.  This approach has elucidated the transition to turbulence when the laminar state remains stable and a boundary in phase space exists between states which excite turbulence and those which decay \citep{Schneider:2007kl,Pringle:2012cb}. In sustained turbulence it is hoped that such solutions act as proxies for the complexity of the flow and so help to unravel the processes sustaining turbulence \citep{Hamilton:1995,Kawahara:2012iu,Lucas:2017fz}. There is also a hope that ECSs can act as a reduced description of the chaos and periodic orbit theory can be used to reconstruct turbulent averages \citep{Chandler,Cvitanovic:2013fz}.
 
Despite these successes the computational methods used so far have some important shortcomings. The current state of the art for converging unstable periodic orbits (UPOs) form what has become known as the ``recurrent flow analysis',' pioneered in \cite{Kawahara:2001ft,Viswanath:2007wc, Cvitanovic:2010, Chandler}. This requires near recurrent episodes to be located in numerical simulations which form guesses for a high-dimensional Newton solution of the recurrence condition $\bm{x}(t)-\bm{x}(t-T)=0,$ for an orbit of period $T$. The algorithms circumvent the formation of the Jacobian matrix by way of a GMRES solution (or similar Krylov method) and maintain the Newton step-size within a trust region of its linearisation by a hookstep \citep{Viswanath:2007wc}. We refer to this solution algorithm as Newton-GMRES-hookstep (NGh). Such algorithms, as with any Newton method, require an initial guess sufficiently close to the solution for guaranteed convergence. This becomes increasingly difficult to determine for more severe turbulence where instability is increased and close approaches to a target solution are more fleeting. Moreover the basins of attraction for convergence are highly complex and usually fractal in nature, and therefore convergence is very difficult to predict. By far the biggest computational inefficiency with the recurrent flow analysis is in the resource spent attempting convergences which fail or result in a known ECS. 
There is significant room for improvement and several subsequent studies have been working on refinements or alternatives, including using dynamic-mode-decomposition \citep{page_kerswell_2020}, variational methods \citep{Lan:2004ch, azimi2020} and preconditioning \citep{tuckerman_langham_willis_2018}. {A promising approach which obtains solutions with a single unstable direction embedded in the `edge' manifold by a feedback control method has recently shown significant efficiency savings by avoiding costly bisection iterations \cite{willis_2017}.} One objective of this paper is to trial a more general control method which can stabilise solutions with high dimensional unstable manifolds so that they may be obtained simply by time-stepping a slightly modified set of equations.

For small systems of nonlinear ordinary differential equations a method known as time-delayed feedback control (TDF) attributed to Pyragas \citep{Pyragas:1992ch} (and so is also known as Pyragas control in the literature) has seen considerable success at stabilising periodic orbits from chaotic systems. The key idea is to include into an evolution equation of the form $\dot{\bm x}=\bm f(\bm x),$ an additional time delayed difference term;
$$\dot{\bm x}=\bm f(\bm x) +  G(t) \left(\bm x(t-T) - \bm x(t)\right).$$
It can be shown that for a given period $T$ and {gain} matrix $G$ this additional delay difference term can stabilise some periodic orbits. Notice that this term has the property that for a periodic solution with period $T$ it vanishes identically. This means that such a solution of the controlled system is also a solution of the original system. The method is therefore termed `noninvasive'. The method is particularly appealing because of its simplicity, any direct numerical simulation code can be easily adapted to include the extra terms, and it does not require \emph{a priori} knowledge of the controlled solution, as for other control methods \citep{Pausch_2011,Smaoui_2017,linkmann_2020}. Only the delay period is required, which it has been shown can be iteratively obtained \citep{herrmann2001, Shaabani2017}.

This method, and its variants have seen success in a variety of systems, for example semiconductor lasers \citep{Ushakov:2004em,Schikora:2006dj}, neuroscience \citep{Popovych:2005hj,Scholl:2009}, microscopy \citep{Yamasue:2006} and chemical turbulence \citep{Kim:2001}. Delayed feedback has also been used to stabilise standing waves in complex Ginzburg-Landau equations \citep{Stich:2013}, and an experimental study controlling Taylor-Couette flow \citep{Luthje:2001do}. 
To the best of our knowledge only two applications of the method for the Navier-Stokes equations have been reported \cite{Kawahara2005,Shaabani2017}. In \cite{Kawahara2005} Kawahara reports the result of stabilising the gentle periodic orbit of \cite{Kawahara:2001ft} using Pyragas control, while this is highly encouraging there is not much guidance on how one may effectively employ TDF in the fluid problem. 
More recently Shaabani et. al. \cite{Shaabani2017} report the application of Pyragas control to suppress vortex pairing in a periodically forced jet. This work approaches the control method as a way of filtering out non-harmonic frequencies, leaving only $T$ behind. These authors report a number of interesting results, including reducing the memory burden of storing the history vector by interpolation between checkpoints, as well as the application of a method to converge $T$ when it is not known \emph{a priori.}  These studies serve as good motivation for a systematic attempt at using the method to stabilise multiple nonlinear solutions embedded in the chaotic set.

One sticking point of TDF is that it is argued that orbits with {an odd number of real, positive} unstable Floquet multipliers are unable to be stabilised by this method \citep{Just1999,  Nakajima:1998kw}. 
{An explanation of this feature is provided in \citep{Nakajima1997} from the perspective of bifurcation theory. First suppose a UPO is stabilised at a certain $G=\bar{G},$ and delay period $T.$ This means between $G=0$ (the uncontrolled system) and $G=\bar{G}$ there is a change in stability of the orbit, and therefore a bifurcation. Next, as mentioned earlier, the number of UPOs of period $T$ cannot vary with $G:$ e.g. an orbit in the controlled system must be an orbit in the original system. This excludes any bifurcations which involve a change in the number of period-$T$ orbits through coalescence, for example pitchfork or saddle-node (see later for a transcritical exception). Any bifurcation must therefore involve complex Floquet exponents crossing the imaginary axis (Hopf or period-doubling), requiring an even number of exponents to form conjugate pairs. It should be noted that there have since been various studies offering resolutions to this issue, including forcing oscillation of the unstable manifold through $G$ \citep{Schuster1997,Flunkert:2011cg} and by counter example \citep{Fiedler:2011db,Sieber:2016cu}. Fiedler et. al. \citep{Fiedler:2011db} show that an orbit with a single unstable Floquet multiplier can be stabilised through a transcritical bifurcation, using complex gain, where the exchange of stability occurs with a delay-induced orbit which only has a period matching the target UPO at the point of bifurcation, thereby avoiding the issues described by Nakajima \citep{Nakajima1997}. Generically we must be aware of this limitation when tackling UPOs in the Navier-Stokes equations. From the perspective of the method as a frequency damping technique, as described in \citep{Shaabani2017}, one can view this limitation as restricting us to stabilising orbits which have only oscillatory unstable manifolds (at least locally); where an unstable direction is ``torsion free'' there is no incipient frequency to damp and this growth can go unchecked by TDF. We should also emphasise that the odd-number limitation is a condition under which we should expect TDF to fail; there is no guarantee TDF will succeed should we have a solution with an even number of unstable eigenvalues.}

In this paper we seek to address several outstanding questions regarding the application of TDF to the Navier-Stokes equations, namely can ECSs be stabilised and is there a simple way to avoid the `odd-number' limitation? If so, what are the requirements for success, and how can we develop the method into a practical tool for the dynamical systems approach to turbulence?

The paper is organised as follows. Section \ref{sec:form} describes the system under consideration and the methods used. Section \ref{sec:lam} shows a linear stability analysis for the basic flow and validates the numerical application of TDF by stabilising the laminar state in a direct numerical simulation (DNS) and provides insight for the effective application of the method. Section \ref{sec:TW} demonstrates the stabilisation of travelling waves using an adaptive method to fix the phase speed. We demonstrate the work-around of the odd-number limitation which arises naturally at high Reynolds numbers by applying an additional symmetry operation in the feedback term. In section \ref{sec:sym1}, by exploring different possible symmetry combinations and multiple delay terms, we discover that eight more equilibria and travelling waves can be stabilised with a single parameter set.  
Finally in section \ref{sec:disc} we summarise and discuss the results before considering possible avenues of further work.

\section{Formulation}\label{sec:form}
In this paper we will present the application of time-delayed feedback control to Kolmogorov flow; the sinusoidally body forced incompressible two-dimensional Navier-Stokes equations. This flow is widely studied both for  transition to turbulence and for the recurrent flow analysis mentioned in the introduction \citep{Chandler, Lucas:2015gt}. We consider a vorticity formulation for which the equations, in nondimensional form, are

\begin{align}
\frac{\partial \omega}{\partial t} + \bm{u}\cdot\nabla \omega &= \frac{1}{Re} \Delta \omega -n \cos (n y) + f.\label{eq:NS}\\
\nabla\cdot \bm u &=0 \label{eq:incom}
\end{align}
with vorticity $\omega = \nabla \times \bm{u} \cdot \hat{\bm z},$ $(\hat{\bm x},\,\hat{\bm y},\,\hat{\bm z})$ being the standard cartesian unit vectors, velocity $\bm{u},$ $Re$ the Reynolds number and $ f $ is a second forcing term. We will consider the periodic torus $[0,2\pi]\times[0,2\pi]$ and a forcing wavenumber $n=4$ and solve the equations with a standard pseudospectral method using two-thirds dealiasing, fourth order Runge-Kutta timestepping on the nonlinear and forcing terms and Crank-Nicolson on the viscous term. For $Re\leq 40$ a resolution of $128^2$ is used and $200\geq Re>40,$ $256^2.$ The code is implemented in CUDA to run on GPUs and available at \url{https://bitbucket.org/dan_lucas/PSGPU} with a Python version in Jupyter notebooks available in supplementary material \cite{SupCode}.

The standard, $f=0,$ Kolmogorov flow system is invariant under the symmetries

\begin{align}\label{eq:sym}
\mathcal{S}:[u,v,\omega](x,y) &\rightarrow [-u,v,-\omega]\left(-x,y+\frac{\pi}{n}\right),\\
\mathcal{R}:[u,v,\omega](x,y) &\rightarrow [-u,-v,\omega]\left(-x,-y\right),\\
\mathcal{T}_s:[u,v,\omega](x,y) &\rightarrow [u,v,\omega]\left(x-s,y\right) \qquad \textrm{for } 0\leq s \leq {2\pi},
\end{align}

\noindent
where $\mathcal{S}$ represents the discrete shift-and-reflect symmetry, $\mathcal{R}$ rotation through $\pi$ and $\mathcal{T}_s$ is the set of continuous translations by $s$ in $x$.

\subsection{Flow measures}

In order to discuss various features of the flows considered we define here some diagnostic quantities. Total energy, energy dissipation rate and energy input rate are defined in the standard way as
\begin{align}
E(t) &:= \half \langle \bm u^2 \rangle_V,\qquad
D(t) := \frac{1}{Re} \langle|\nabla \bm u |^2\rangle_V, \qquad
I(t) :=  \langle u \sin(ny) \rangle_V .
\end{align}
where the volume average is defined as 
$ \langle\quad\rangle_V := \frac{1}{4\pi^2} \int_0^{2\pi}\int_0^{2\pi} \ud x \ud y. $ Note the energy budget is such that $\ud E / \ud t = I-D$ meaning that any steady state, and any time average, must satisfy $D=I.$

For Kolmogorov flow the basic flow, which is the global attractor at small Reynolds number $Re,$ is given by the precise balance between forcing and dissipation; the profile and its energy and dissipation rate are 
\begin{align}
\bm u_{lam} := \frac{Re}{n^2}\sin ny \bm{\hat x}, \qquad \omega_{lam} := \frac{Re}{n}\cos ny \qquad E_{lam} := \frac{Re^2}{4n^4}, \qquad D_{lam} := \frac{Re}{2n^2}.
\label{eq:base}
\end{align}
\subsection{Time-delayed feedback}
Time-delayed feedback is included by setting 

\beq
f = G(t)\left(\psi(\bm{x},t)-\psi(\bm{x},t-T)\right)
\label{eq:f}
\eeq
where the streamfunction, $\psi,$ is defined as $\bm u = (\psi_y,-\psi_x),$ such that $\omega = -\nabla^2 \psi.$ $T$ is the delay period, $G(t)$ is a scalar gain function.

It should be noted that, in principle, we have a great deal of freedom in choosing $G.$ It could be a function of space as well as time, or even be an operator. In fact the choice above is equivalent to $f =-\hat{G}(t)\left(\omega(\bm{x},t)-\omega(\bm{x},t-T)\right)$ with $\hat{G}(t) = G(t)\nabla^{-2}.$ This choice was made in small part to improve the performance of the method; it means that the perturbing feedback force is relatively stronger on large scales ($\nabla^2$ being a factor $-|\bm{k}|^2$ in Fourier space)  compared to $f=G(t)\left(\omega(\bm{x},t-T)-\omega(\bm{x},t)\right)$. It is mainly motivated by allowing a cleaner linear analysis in the next section. 

While the above form of $f$ is the basic delay difference for TDF, we will show that exploiting the symmetries of unstable solutions can improve the ability of TDF to stabilise them. Moreover, to obtain travelling wave solutions or relative periodic orbits, we must translate the delayed term by a distance $s$ at the rate dictated by the phase speed of the solution $c = s/T.$ To this end we will include a translation symmetry transformation \eqref{eq:sym} to one of the terms in the TDF forcing such that 

\beq
f = G(t)\left(\psi(\bm{x},t)-\mathcal{T}_s\psi(\bm{x},t-T)\right), \label{eq:TDF_f}
\eeq
or equivalently
\beq
f = G(t)\left(\psi(\bm{x},t)-\psi(\bm{x}-s\hat{\bm x},t-T)\right).
\eeq
We will consider including the shift-and-reflect and rotational symmetries in later sections.

\section{Linear stability analysis of the basic flow}\label{sec:lam}

In order to give some theoretical motivation and validation of the TDF method we first apply it to the basic flow, equation \eqref{eq:base}.
This laminar state has a well-known linear instability \citep{Meshalkin:1961ty} at a critical Reynolds number, $Re_c.$ For $n=4$ this is $Re_c\approx 9.97.$ The linear analysis can be extended to approximately include the time-delayed term. Rewriting Equation \ref{eq:NS} in terms of streamfunction, expanding about the base flow, $\psi = \psi_{lam} + \psi',$ and linearising yields

\begin{align}\label{psi_linear}
\frac{\partial \nabla^2\psi'}{\partial t} + Re\sin(ny)\left(\frac{1}{n^2}\nabla^2\psi_x' + \psi_x'\right) &= \frac{1}{Re} \nabla^4\psi' + G\left(\psi'(\bm{x},t)-\psi'(\bm{x}-s\hat{\bm x},t-T)\right)& \end{align}
using the usual ansatz $\psi' = \hat{\psi}(y)e^{\irm\alpha x + \sigma t}$ the modified Orr-Sommerfeld equation now reads

\begin{align}
\sigma\left(\frac{\ud^2 }{\ud y^2}-\alpha^2\right)\hat{\psi} =&  - \irm\alpha Re\sin(ny)\left(\frac{1}{n^2}\left(\frac{\ud^2 }{\ud y^2}-\alpha^2\right)\hat{\psi} + \hat{\psi}\right) \nonumber \\ & + \frac{1}{Re}\left(\frac{\ud^2 }{\ud y^2} -\alpha^2\right)^2\hat{\psi}  +  G\left(1 - e^{-\irm\alpha s - \sigma T }\right)\hat{\psi}
\end{align}
The equation is now transcendental in the eigenvalue $\sigma$ due to the exponential coming from the delay term. If we assume $T$ is small but nonzero (reasonable since we are attempting to stabilise the laminar equilibrium and not a UPO) 
and are only interested in $\sigma \approx 0,$ (again reasonable if we are tracing stability boundaries and not interested in growth rates in general) then we may expand the exponential to linear terms in $\sigma,$ i.e. $e^{-\sigma T} \approx 1-\sigma T$ resulting in 

\begin{align}
\sigma\left(\frac{\ud^2 }{\ud y^2}-\alpha^2-GTe^{-\irm\alpha s}\right)\hat{\psi} &=  - \irm\alpha Re\sin(ny)\left(\frac{1}{n^2}\left(\frac{\ud^2 }{\ud y^2}-\alpha^2\right)\hat{\psi} + \hat{\psi}\right) \nonumber  \\ +& \frac{1}{Re}\left(\frac{\ud^2 }{\ud y^2} -\alpha^2\right)^2\hat{\psi}  +  G\left(1 - e^{-\irm\alpha s }\right)\hat{\psi}.\label{eq:OSmod}
\end{align}

\begin{figure}
\begin{centering}
 \includegraphics[width=0.49\linewidth]{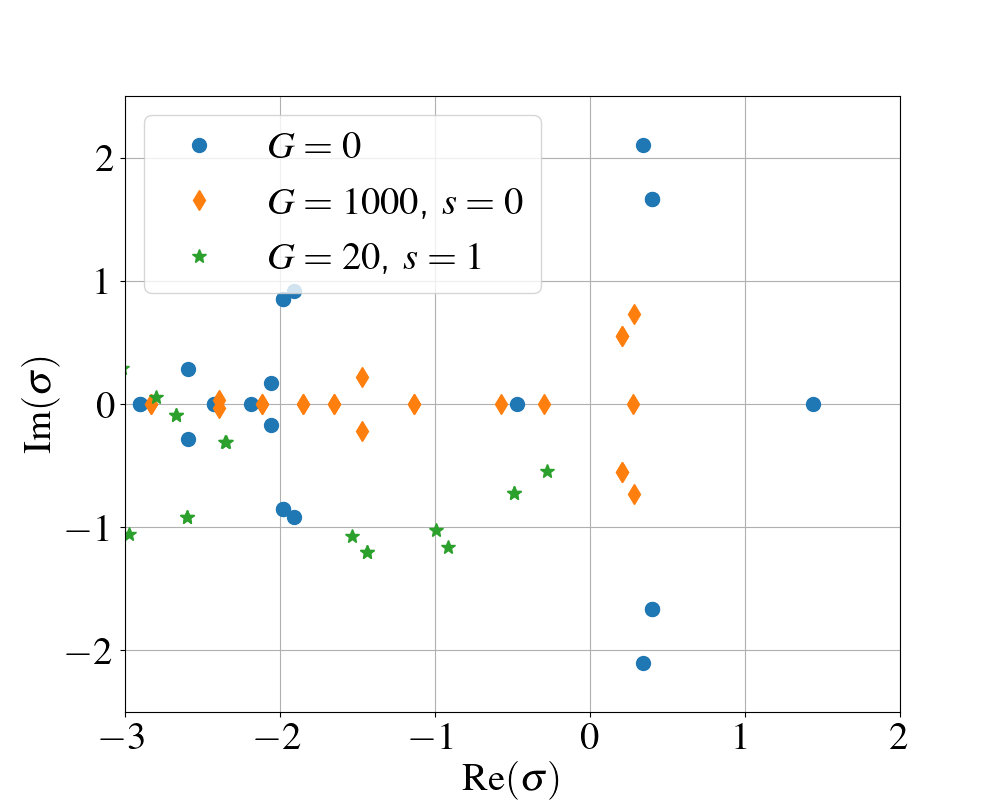}
\includegraphics[width=0.49\linewidth]{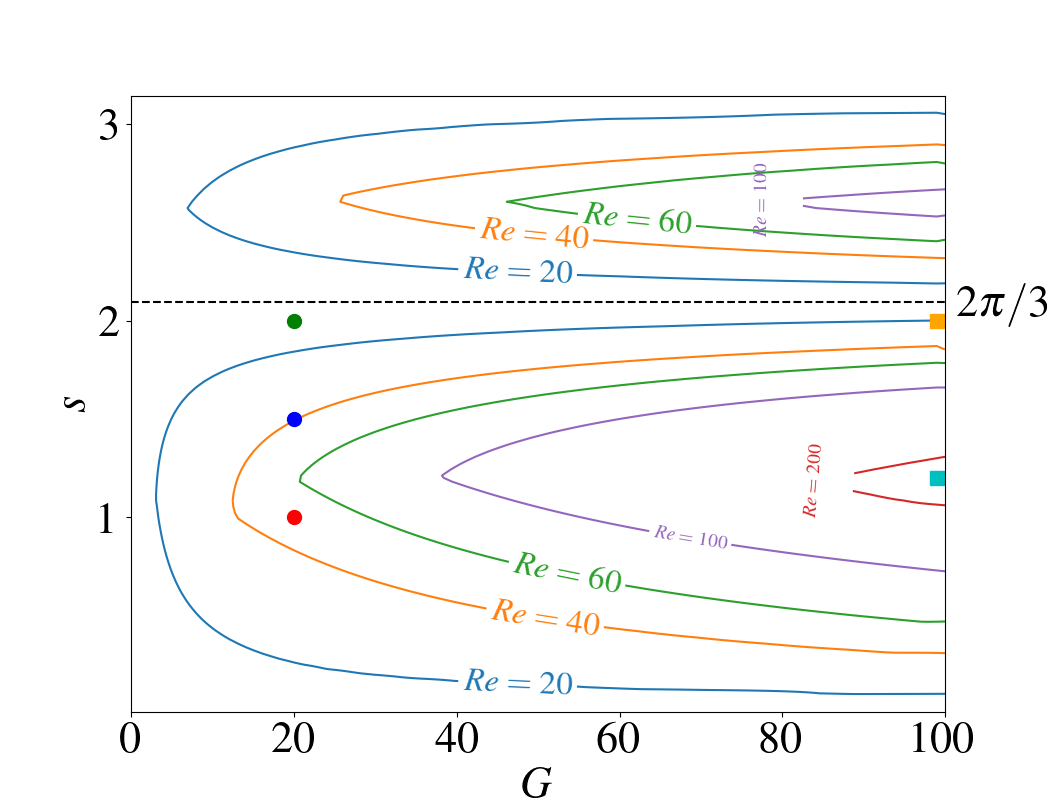}
\caption{\label{fig:eigs} (Left) Eigenvalue spectra for the modified Orr-Sommerfeld operator with $\alpha=1,$ $R=40.$ Blue circles show the result for $G=0,$ i.e. the uncontrolled Kolmogorov flow case. Orange diamonds show the rescaling of the spectrum for $G=1000$ and $s=0.$ Green stars show the shifting of the spectrum for $G=20$ and $s=1$ demonstrating the crossing of the imaginary axis of the eigenvalues with largest real part. (Right) Neutral curves for various Reynolds numbers in the $(s,G)$ plane with stable region lying to the right of the curves. {Contours (outward in) $Re=20$ blue, 40 orange, 60 green, 100 purple, 200 red.} Of note are the regions about $s=\pi\,\&\,\frac{2\pi}{3}$ (dashed line) where instability is always found regardless of the size of $G$ (the curves are generated for all $\alpha$). The symbols are included to indicate where the DNS validation has been conducted (circles $Re=40,\, G=20,$ squares $Re=200,\, G=100$), in particular the blue circle at $G=20,\, s=1.5$ lies on the $Re=40$ neutral curve. 
}
\end{centering}
\end{figure}

We solve the eigenvalue problem numerically using a Fourier series expansion $\hat{\psi}(y)= \sum_k\hat{\Psi}_ke^{\irm ky}.$ The result is a generalised eigenvalue problem of the form
\beq \sigma\bm{B \hat{\Psi}} =  \bm{A \hat{\Psi}} \qquad \Rightarrow \qquad \sigma\bm{ \hat{\Psi}} =  \bm{B^{-1} A \hat{\Psi}}, \label{eq:eig} \eeq
where
\begin{align}
\bm{ \hat{\Psi}} &= \begin{pmatrix} \vdots \\ \hat \psi_{k} \\ \vdots  \end{pmatrix},\qquad B_{ij} = \begin{cases} k^2 + \alpha^2 + GTe^{-\irm\alpha s}, & i=j \\ 
0 & i\neq j, \end{cases} \label{eq:B} \\
A_{ij} &= \begin{cases} -(k^2 + \alpha^2)^2/Re + G(e^{-\irm\alpha s}-1), & i=j, \\ 
-\frac{Re\alpha}{2n^2}(k^2 + \alpha^2 - n^2), & i=j-n, \\
\frac{Re\alpha}{2n^2}(k^2 + \alpha^2 - n^2), & i=j+n, \\
0, & \mathrm{otherwise}. \end{cases}\label{eq:A}
\end{align}
The spanwise wavenumber $k = i-M$ and $N = 2M+1$ is the total number of Fourier modes set to 33 in the results to follow.
Note that $\bm{B}$ is diagonal and therefore trivial to invert, however it is useful to separate the contributions from $\bm{A}$ and $\bm{B}$. We solve the eigenvalue problem in Python using the numpy eigenvalue solver \citep{2020SciPy-NMeth,Strang}. 
Figure \ref{fig:eigs} (left panel) shows the relevant part of the eigenvalue spectrum at $Re=40$ and $\alpha=1$ without any feedback, $G=0.$ This is well above criticality, in fact the flow is chaotic at $Re=40$ and $\alpha=1$ is the first streamwise mode to become unstable. There are five unstable modes with positive real part (one purely real and two complex conjugate pairs). Setting $G=1000,\,\, T=0.01$ but no translation $s=0$ shows the unstable spectrum being rescaled toward the origin. Setting $G=20,$ and $T=0.01$ but now $s=1$ we see the whole spectrum shift/rotate and all of the unstable modes cross the axis, including the purely real ones. 

We can interpret the effect of TDF on the laminar solution by examining the effect of the terms involving $G,$ in equation \eqref{eq:eig} and the matrices \eqref{eq:B} and \eqref{eq:A}. The contribution involving $G$ to the matrix $\bm A$ is homogeneous (in $y$) meaning one can write it as a constant, $g = G(1-e^{-\irm\alpha s}),$ multiplying the identity. Such a transformation, $\bm{C}-g\bm{I},$ will translate the eigenvalue spectrum of $\bm{C}$ by $g.$ When $s=0$, or indeed $\alpha s = 2\pi,$ this translation disappears as $g=0.$

The contribution of the TDF terms to $\bm B$ when $s=0$ does not translate the spectrum, it adjusts the multiplicative effect of $\bm B^{-1}.$ For positive $G$ this will make each entry of $\bm B^{-1}$ smaller (recall $\bm B$ is diagonal) hence rescaling eigenvalues toward the origin, but never crossing the imaginary axis.  For $G<0$ the opposite is true and the spectrum is inflated, however there are critical levels where $GT = -(k^2 + \alpha^2)$ and $\bm B$ becomes singular. If $GT\ll-(k^2 + \alpha^2)$ then a change of sign can occur and previously stable eigenvalues become unstable. This observation is useful as it justifies the sign choice for $G$ in what follows; direct numerical simulations blow up for negative $G.$ If $s\neq 0$ then the `rotation' embedded in $\bm{B^{-1}}$ can also contribute to the stabilisation, however the combined effect of $T$ and $s$ becomes non-trivial. Needless to say if $\sigma T$ is small, as assumed, this rotation will also be a small correction.

Because the main effect of the translation $s$ is at $O(1)$ not $O(T)$ in \eqref{eq:A} it remains as $T \to 0.$ This limit would be equivalent to no time-delay and a perturbing force $f = G(t)\left(\psi(\bm{x},t)-\mathcal{T}_s\psi(\bm{x},t)\right).$ Indeed we find stabilisation of the laminar profile can also be achieved without a delay, provided $s$ is chosen correctly. Although we do not consider this case further for the laminar flow, we will revisit this situation in section \ref{sec:T} when considering nonlinear equilibria. One consequence of these observations is that for a general equilibrium solution with $s=0$ and no other symmetry (rotation or shift-reflect), TDF will always fail. 

{In this implementation we have translated the delayed term forward in $x,$ however we should also address why one should not translate the current state back, in other words setting $f = G(t)\left(\mathcal{T}_{-s}\psi(\bm{x},t)-\psi(\bm{x},t-T)\right)$ rather than equation \eqref{eq:TDF_f}. In the limit $T\to0$ it is clear this choice is equivalent to reversing the sign of $G$ and will destabilise rather than stabilise. One could recover stabilisation by choosing negative $G.$ However for nonzero $T$ and negative $G$ the `critical levels' described above come into play where $\bm{B}$ can become singular. Therefore in this formulation there is a clear and correct choice for how to set the TDF terms as shown in equation \eqref{eq:TDF_f} and ensuring $G>0$.}

To get a broader picture of how this stabilisation depends on $G,$ $s,$ and $Re,$ figure \ref{fig:eigs} (right panel) shows neutral curves on the $(G,s)$ plane for various $Re$ now accounting for all unstable $\alpha,$ {namely $\alpha=1,\,2,\,3$ (note the left panel is for $\alpha=1$ only)}. The stable region is to the right of the contour, i.e. large $G.$ This shows that there is an interval of $s$ that can stabilise the base flow which reduces in size as $Re$ increases, but can be increased on increasing $G.$ There are two branches, one approximately centered at $s=1$ and one around $s=2.5$ with a gap of instability around $s=2$ regardless of $Re$ or $G.$ We can understand the structure by referring back to the shifting of the eigenvalues by $g$ described above; when $\alpha s = 2\pi$ this term vanishes and the main effect of this translation in $x$ on the spectrum is lost. Up to $Re=200$ only $\alpha=1,\,2$ and 3 are unstable in the uncontrolled case, therefore $s\approx \pi\,\&\,\frac{2\pi}{3}$ will fail to stabilise the laminar solution. 

These predictions can be verified by applying the method to the full nonlinear equations. This is achieved using a Crank-Nicolson-type timestepping (average of forward and backward Euler) for numerical stability. 
Note that we need to integrate the equations for at least $T$ time units before the feedback can be applied. Also if feedback is introduced discontinuously during time-integration, for example in one timestep, then a discontinuity will propagate through the solution in time (a well-known issue in delay differential equations \citep{bellen2003}). To mitigate this we introduce $G(t)$ gradually by the following form

\beq G(t) = \min(G_{max},\kappa(t-T_{start})) \label{eq:G}\eeq
where $\kappa$ is some rate, we use $\kappa=100$ for stabilisation of the laminar state, and $T_{start}$ the time at which we introduce TDF with $G_{max}$ the final maximum. {In the interests of brevity, and to highlight the correspondence with the $G$ used in the linear analysis, when $G$ is referred to in the context of the DNS it is equivalent to the late time $G_{max}$ in \eqref{eq:G}.}

\begin{figure}
\begin{centering}
 \includegraphics[width=0.48\linewidth]{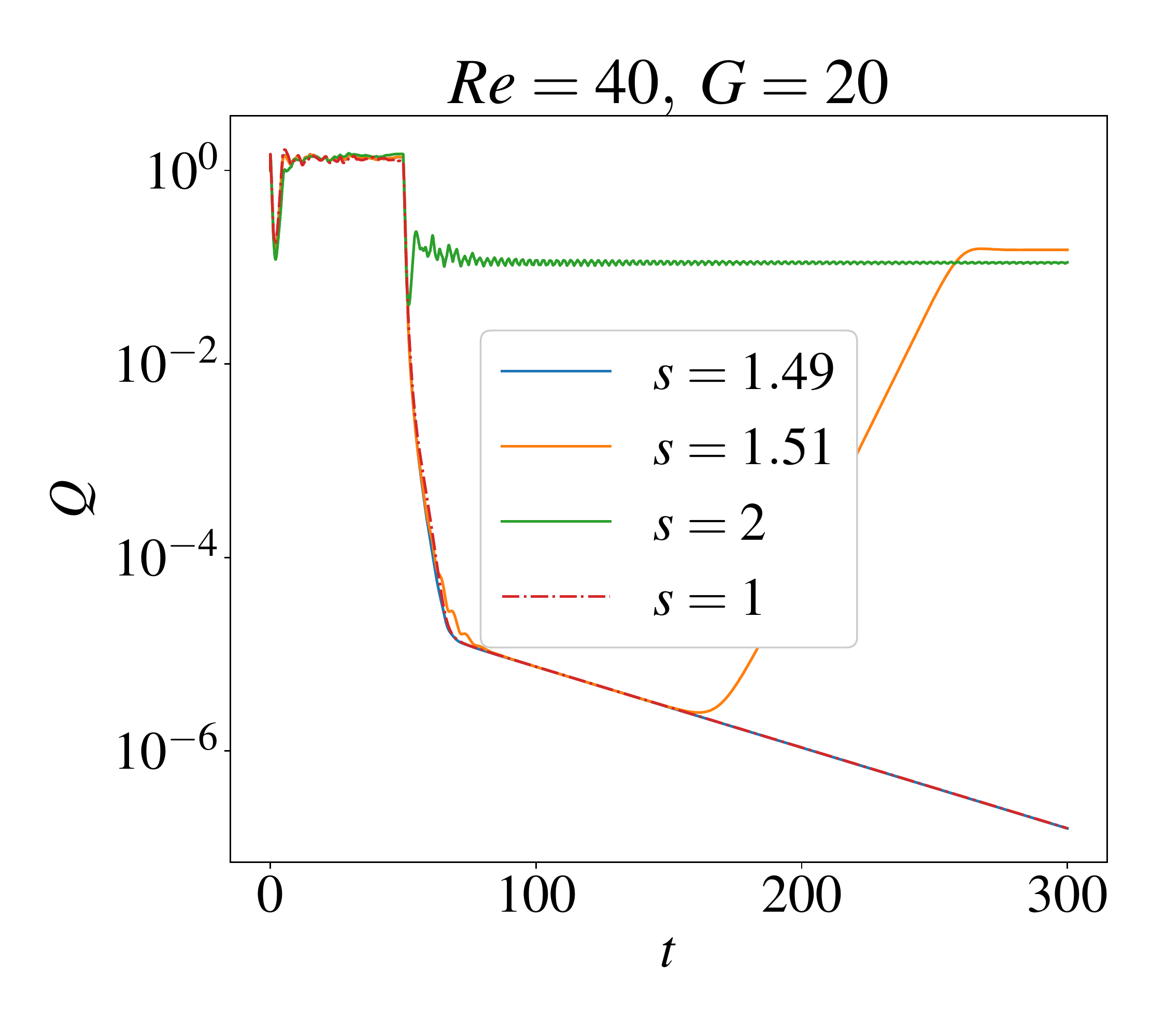}
 \includegraphics[width=0.48\linewidth]{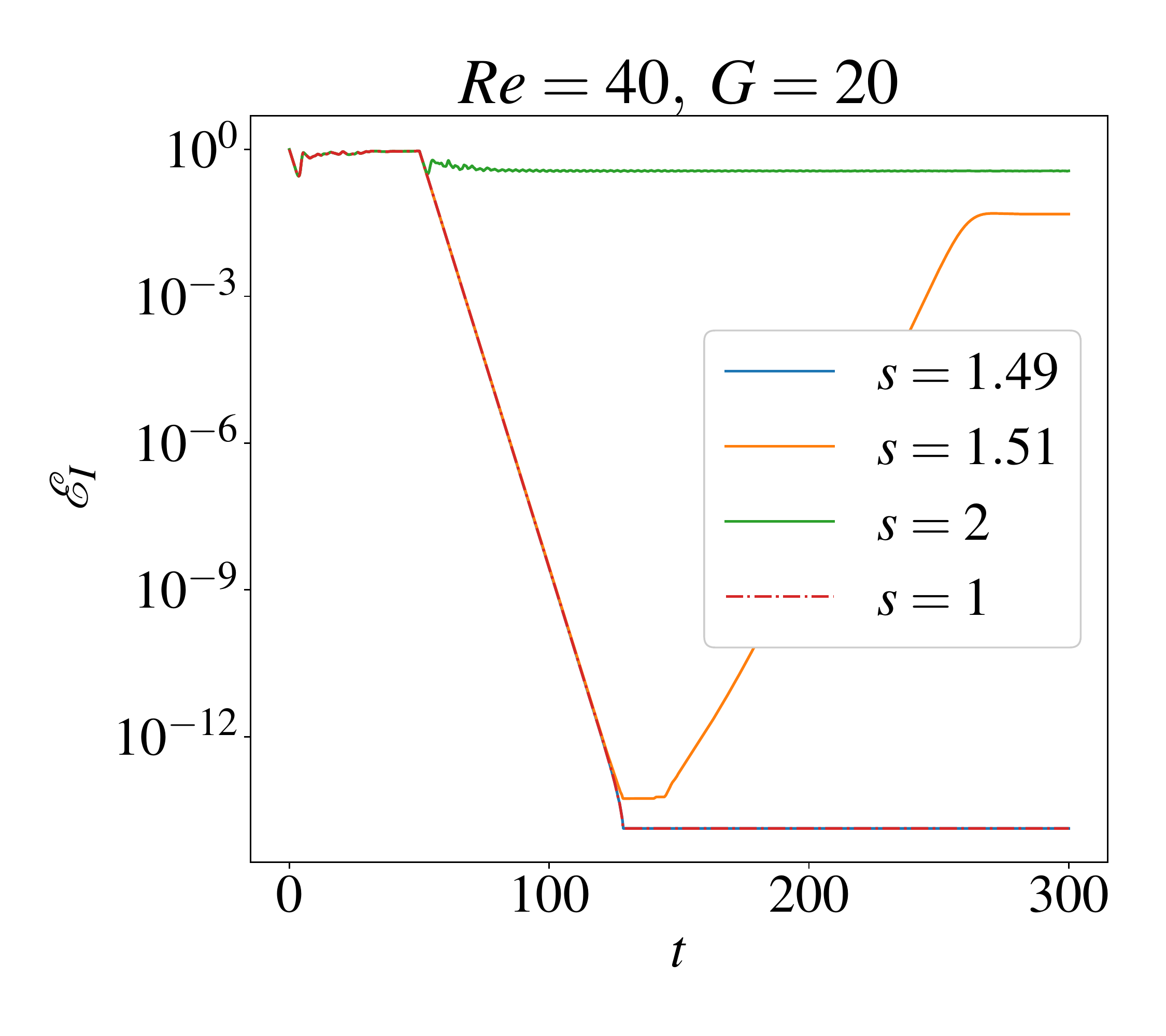}\\
 \includegraphics[width=0.48\linewidth]{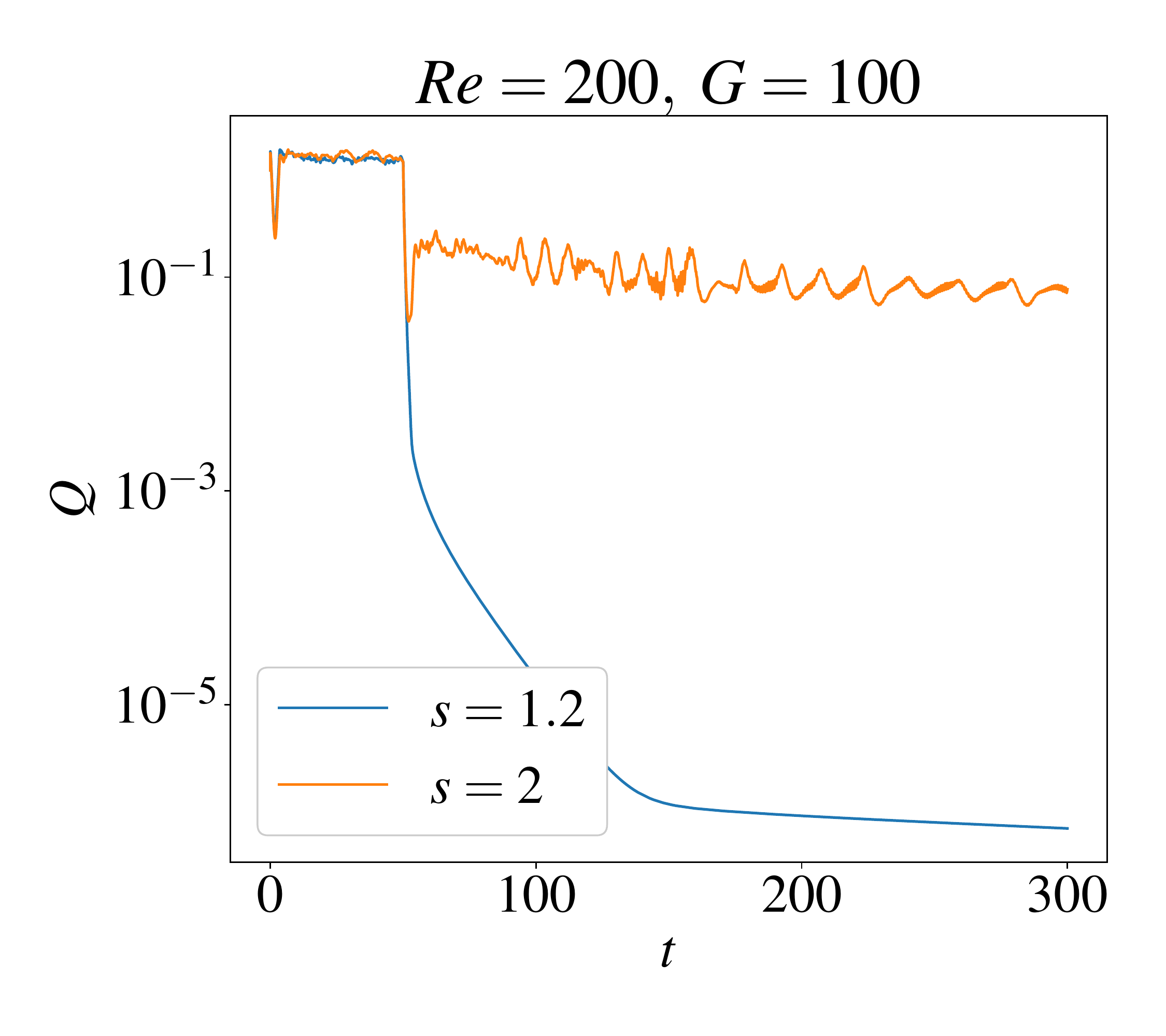}
 \includegraphics[width=0.48\linewidth]{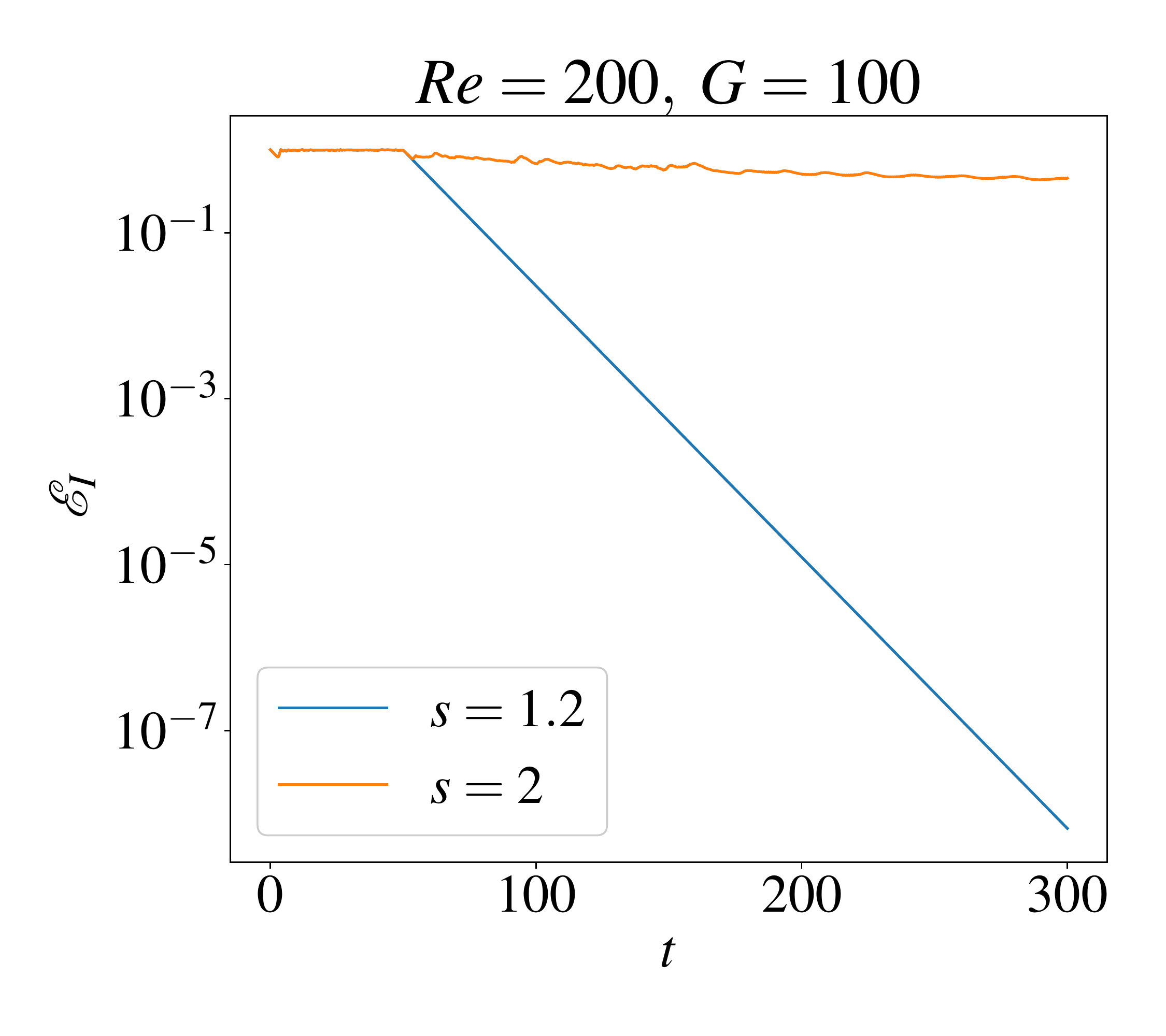}
\caption{\label{fig:lam} {TDF stabilisation of laminar} Kolmogorov flow. Left panels show the relative size of the feedback term $Q$ and right the relative error of the energy input rate relative to the laminar state $\mathcal{E}_I,$  Top row for $Re=40$ and $G=20$ with various choices for $s$ demonstrating stabilisation only within the boundaries shown in figure \ref{fig:eigs}, bottom row at $Re=200,$ $G=100$ with $s=1.2$ showing stabilisation and $s=2$ not, again in agreement with the linear analysis.
}
\end{centering}
\end{figure}

We demonstrate the stabilisation of the laminar solution at $Re=40$ and $Re=200$ in figure \ref{fig:lam} using $T_{start}=50$ to allow ``spin-up'' of the uncontrolled dynamics (from the usual uniform amplitude, $\langle \omega^2\rangle_V=1$, randomised phase initial condition in Fourier space) and $T=0.01$ to be a small delay to give good agreement with the linear theory. To quantify the size of the delay term and approach to the laminar state we introduce the following distance and error measures

\begin{align}
Q(t) &= \frac{\left<\left(\psi - \psi(x-s,y,t-T)\right)^2\right>_V^{1\over 2}}{\left<\psi^2\right>^{1\over 2}_V},
\qquad \mathrm{and} \qquad
\mathcal{E}_I(t) &= \left| 1 - \frac{I}{D_{lam}}\right|.
\end{align}

To demonstrate the precision of the neutral curve estimate we perform numerical simulations with $s=1.49$ and $1.51,$ and also $s=1$ and $2$ to show behaviour well within the stable and unstable regions. Figure \ref{fig:lam} shows that $s=1.49$ gives asymptotic stability of the laminar flow while $s=1.51$ is initially attracted towards $\bm{u}_{lam}$ but then picks up the unstable manifold and moves away. Figure \ref{fig:eigs} indicates at $Re=40$ and $G=20$ a neutral curve is found at $s=1.5,$ marked with the blue circle on the orange $Re=40$ contour. In contrast, far from the neutral curve, $s=2$ does not exhibit a close approach to the laminar flow but settles onto a different attractor in the controlled system. Also of interest is that the curves for $s=1$ and $s=1.49$ are indistinguishable indicating that the decay rate is not sensitive to the specific choice of stable $s.$ The rightmost panel of figure \ref{fig:lam} shows the result of the same type of calculation but at $Re=200$ and $G=100$ with $s=1.2$ and $s=2,$ chosen to lie in the middle of the stable and unstable regions. As predicted $s=1.2$ shows stabilisation and $s=2$ does not. 

\begin{figure}
\begin{centering}
 \includegraphics[width=0.32\linewidth]{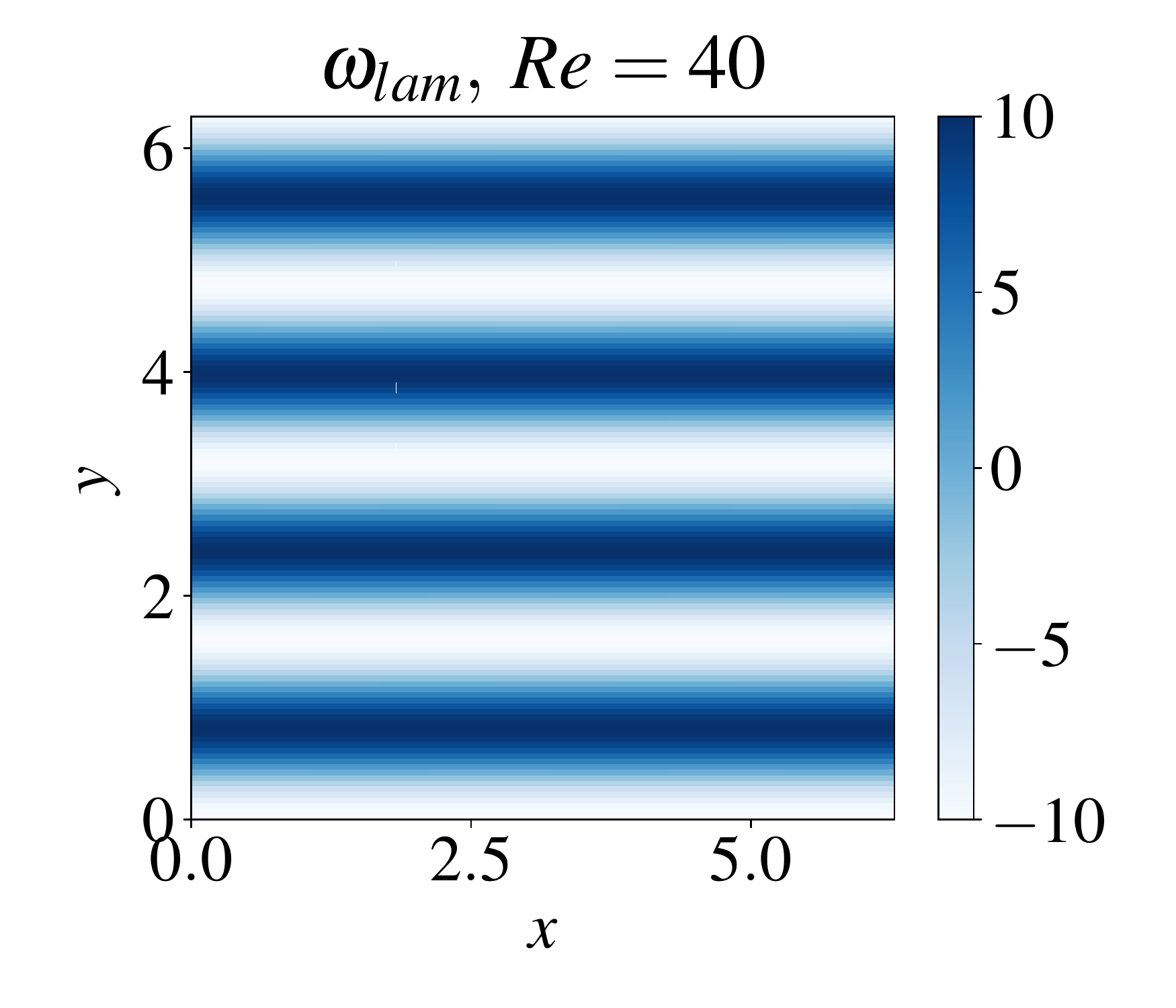}
 \includegraphics[width=0.32\linewidth]{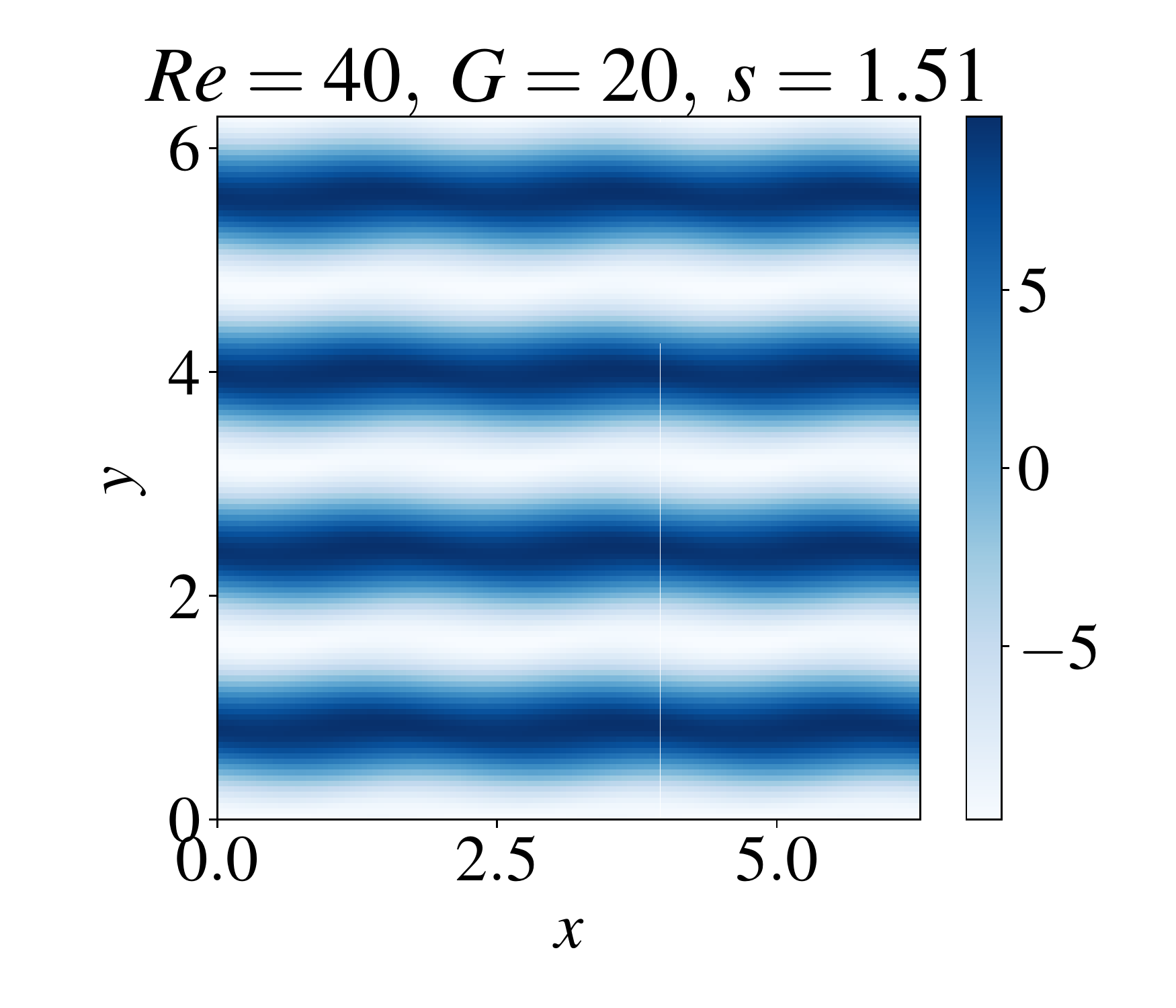}
\includegraphics[width=0.32\linewidth]{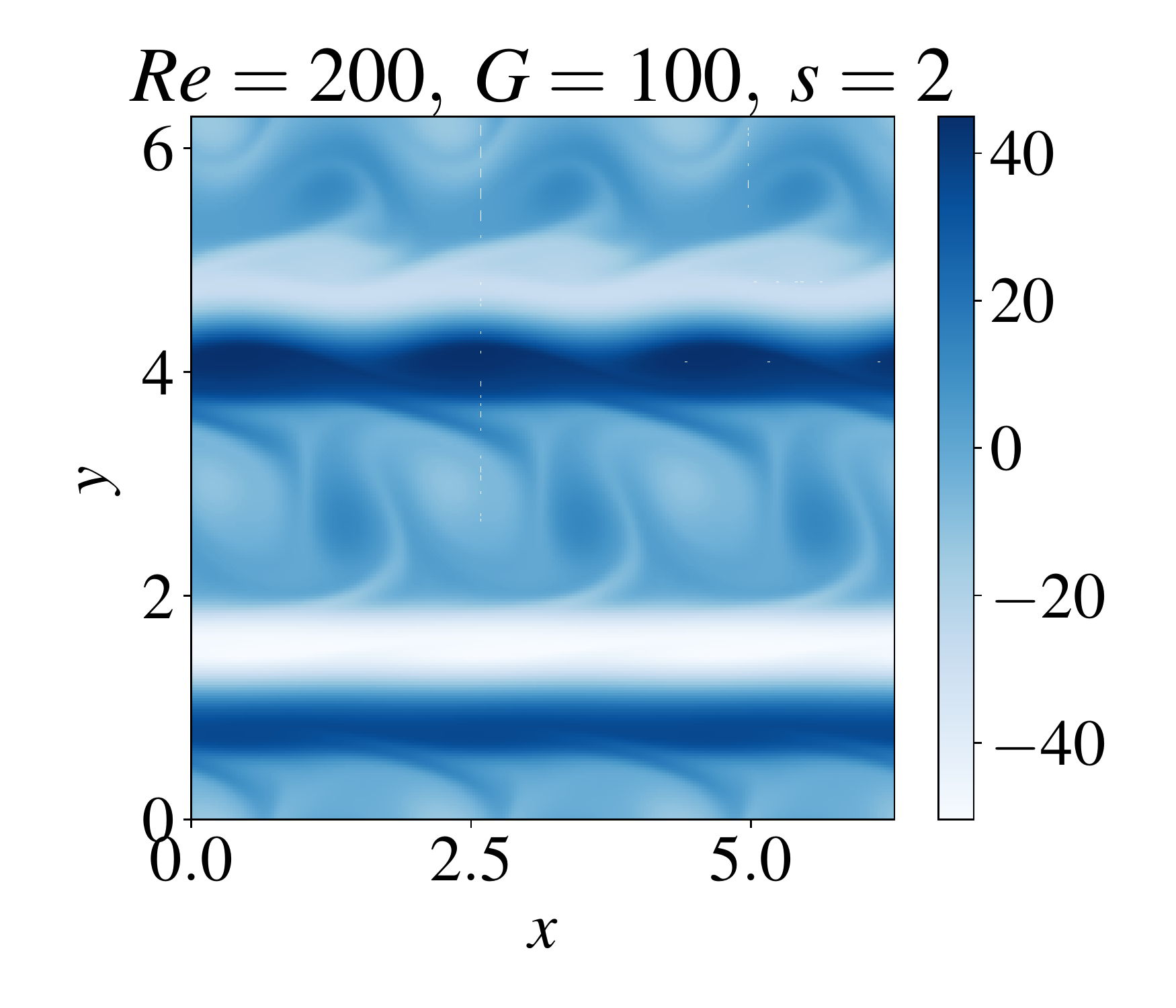}
\caption{\label{fig:vort}Snapshots of the vorticity fields $\omega$ at the end of the simulations shown in figure \ref{fig:lam}. Left shows the stabilised laminar solution at $Re=40,\,\, n=4;$ note the laminar looks exactly the same at $Re=200$ only with larger amplitude. Middle shows the unstable case near the neutral curve $Re=40,\, G=20$ and $s=1.51,$ the state is close to the laminar solution with a streamwise mode three ($\alpha=3$) disturbance in agreement with the linear theory. Right shows the unstable case at $Re=200, \, G=100$ and $s=2,$ the flow field is turbulent but again retains the mode three signature expected when applying TDF for $s\approx\frac{2\pi}{3}$ in this system. 
}
\end{centering}
\end{figure}

Figure \ref{fig:vort} shows snapshots of the vorticity field at the end of these simulations, showing the laminar profile at $Re=40,$ and the flow for $Re=40,\, G=20,\, s=1.51$ and $Re=200,\, G=100,\, s=2.$ We note that in these latter two cases that do not stabilise we see a clear streamwise mode 3 pattern. This is consistent with the analysis that this instability region is due to $\alpha=3;$ the other wavelengths are stabilised and mode 3 persists in the controlled dynamics, even at high $Re$ and large amplitude.



\section{Travelling waves: adaptive phase speed}\label{sec:TW}

The goal of this work is not to merely control turbulence; there is a vast literature on this topic and potentially more effective or applicable methods than TDF. Rather we seek to use this method as an efficient means to discover unstable nonlinear solutions embedded in the turbulent attractor. In this case it is impossible to perform any linear stability analysis \emph{a priori} and predict what parameter values TDF will work for. We will also need to guess the value of the period for UPOs, or phase speed of travelling waves, or indeed translations for relative UPOs. Moreover even if a solution is stabilised there is no guarantee that it will be the unique attractor and that our initial condition will be in its basin of attraction. In general we may desire some adaptive approaches to automatically obtain gain $G,$ period $T$ and shifts $s$ to stabilise target ECSs. 

\begin{figure}
\begin{centering}
 \includegraphics[width=\linewidth]{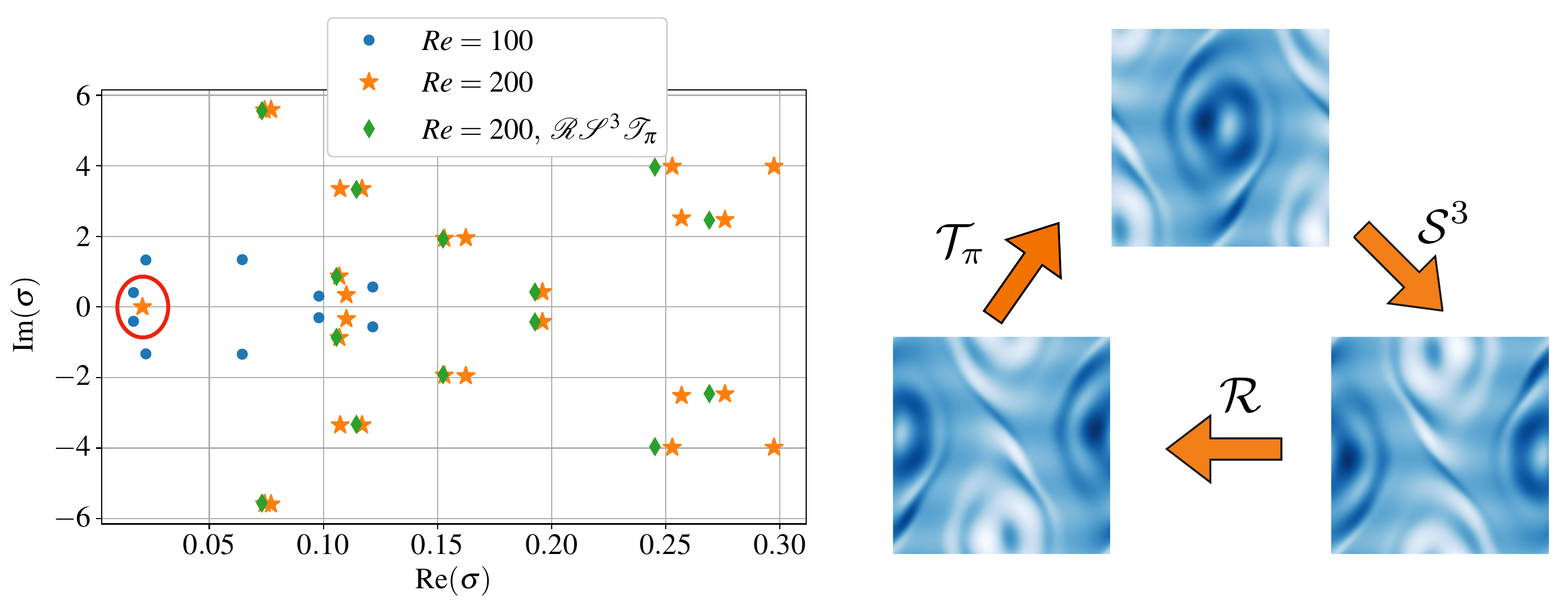}
\caption{\label{fig:TW_eig} Left shows the unstable part of the eigenvalue spectrum  for the $TWa$ travelling wave at $Re=100$ (blue circles) and 200 (orange stars). Note at $Re=200$ the smallest eigenvalue sitting on the real axis circled in red. On projection onto the $\mathcal{R}\mathcal{S}^3\mathcal{T}_\pi$ subspace the solution loses this purely real eigenvalue, this spectrum is shown in green diamonds. Right demonstrates the symmetry operation under which $TWa$ is invariant.  
}
\end{centering}
\end{figure}

As a first step in this direction we demonstrate the case of stabilising the unstable travelling wave which we will denote $TWa,$ reported in \citep{Chandler} as $T1$ for $40\leq Re \leq100$. This solution is  particularly {amenable to our approach} due to being relatively weakly unstable (compared to other solutions) with an unstable spectrum with entirely non-zero imaginary parts (see figure \ref{fig:TW_eig} later). Nonlinear travelling waves, including $TWa$, will 
require that the specific combination for the phase speed of the solution $c=\frac{s}{T}$ to be respected in order for the TDF terms to vanish. While phase speeds for certain travelling waves are reported in the literature, we treat it as an unknown to be computed.

Our approach to finding $c$ is to implement an adaptive method, varying $s(t)$ via gradient descent using a simple ordinary differential equation
\beq \dot{s} = \gamma \delta s \label{eq:s} \eeq
where $\gamma$ is some parameter varying the speed of the descent and $\delta s$ is the translation which minimises the delay difference term $\| \omega(x,y,t) - \omega(x-s-\delta s,y,t-T) \|.$ In other words $\delta s$ is an estimate of the translation remaining between the current flow field and the delayed and translated state. This mean translation is computed by averaging the phase shifts across the individual complex Fourier amplitudes, i.e.
$$s_{\bm k} = \frac{1}{ik_x} \arg \left(\frac{\hat{\omega}_{\bm k}(t)}{\hat{\omega}_{\bm k}(t-T)}\right),$$
where $\omega = \sum_{\bm k} \hat{\omega}_{\bm k} e^{\irm\bm{k}\cdot \bm{x}},$ i.e. $\hat{\omega}$ are the complex Fourier coefficients with $\bm{k}=(k_x,k_y)$ the wavevector, as used in the numerical solution. The individual $s_{\bm{k}}$ being the `shift' required for the phase of that particular mode $\hat\omega_{\bm k}(t)$ to equal $\hat{\omega}_{\bm k}(t-T).$ 
Obtaining these $s_{\bm{k}}$ requires some care with the branches of the complex logarithm (or $\arctan$) when computing the complex argument; the full code is provided in the supplementary material \cite{SupCode}. Once satisfactory $s_{\bm{k}}$ are obtained they are simply averaged such that $\delta s = \frac{1}{N}\sum_{\bm k}^N s_{\bm{k}}-s$ where $N$ will be the total number of dealiased modes in the Galerkin truncation. The ODE is solved alongside the DNS using Adams-Bashforth time-stepping. {Note $\delta s$ could equally be obtained by minimising $\| \omega(x,y,t) - \omega(x-\delta s,y,t-T) \|$ over all translations $\delta s,$ however this would require iteration of some kind, at the very least a trial of a discrete set of $\delta s.$ The method described above is direct as given two state vectors, one for $\omega(x,y,t)$ and one for $\omega(x-l,y,t)$ then $\delta s = l,$ without any iteration and accumulating only rounding errors. }

\begin{figure}
\begin{centering}
 \includegraphics[width=0.32\linewidth]{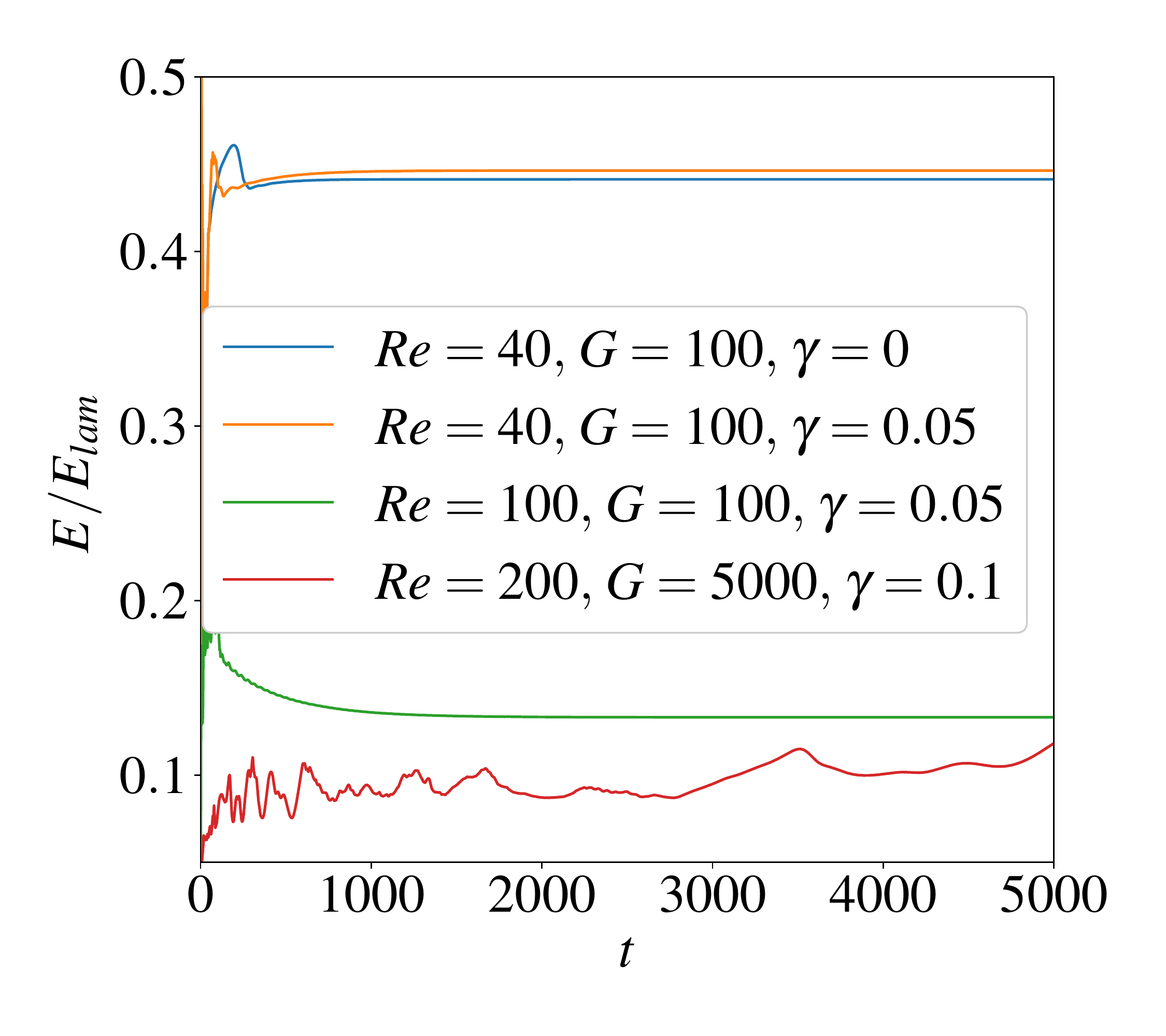}
 \includegraphics[width=0.32\linewidth]{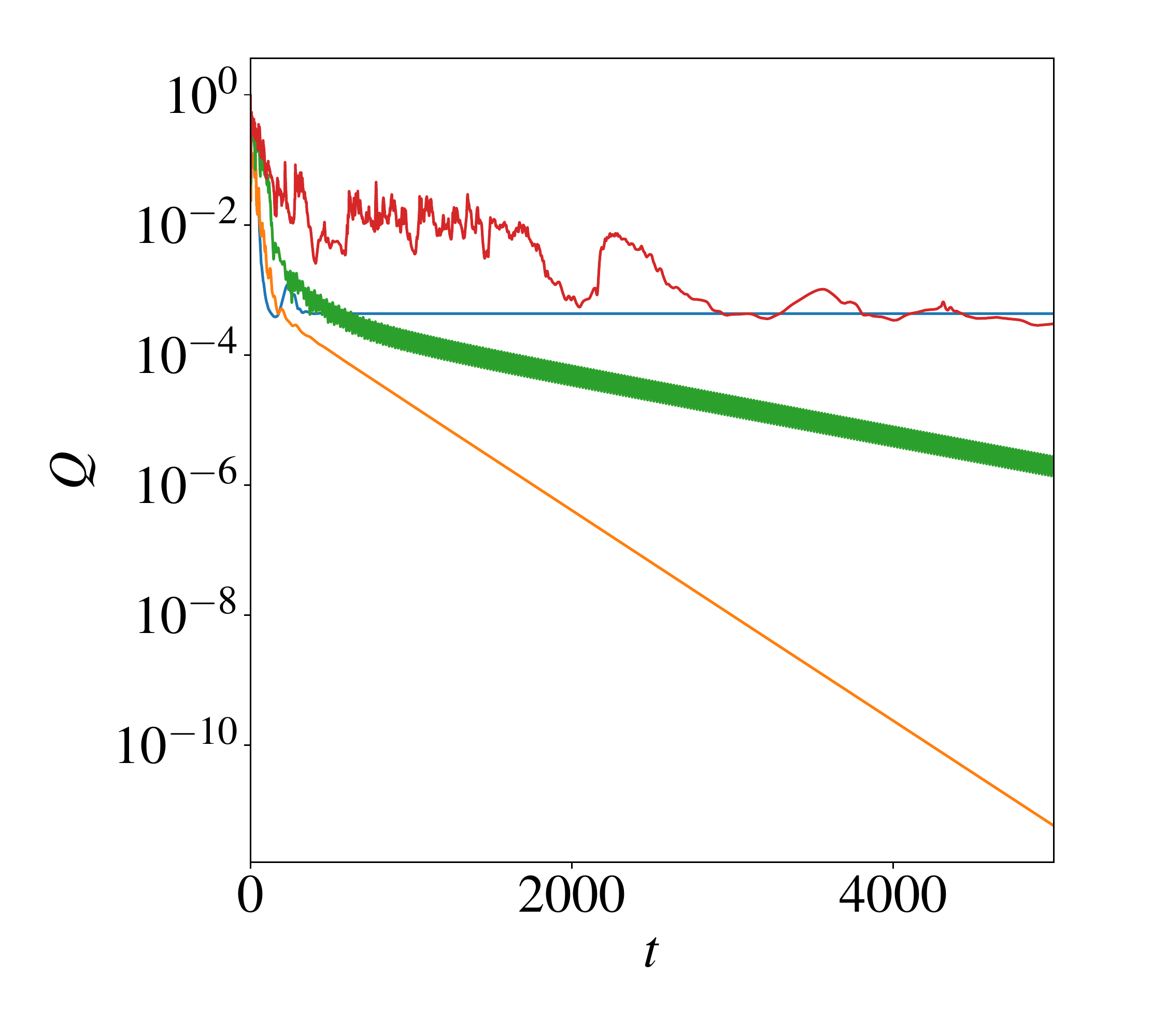}
 \includegraphics[width=0.32\linewidth]{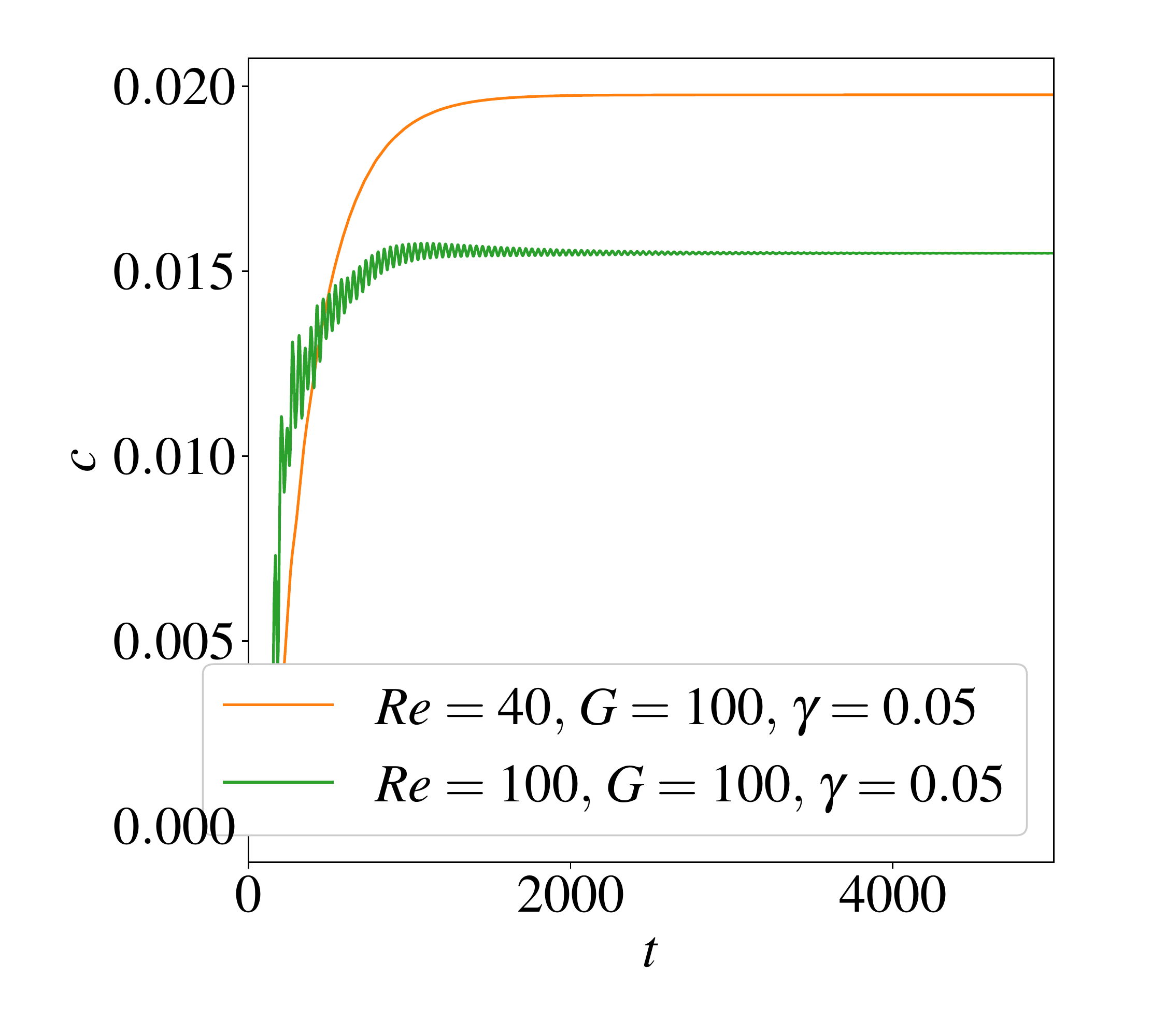}
\caption{\label{fig:TW} TDF stabilisation of the $TWa$ travelling-wave solution of Kolmogorov flow shown via time series of $E/E_{lam},$ $Q$ and $s$ (left to right) when applying TDF at $Re=40$ and $Re=100$ both with $G=100$ and $Re=200$ with $G=5000.$ For $Re=40$ and $Re=100$ the $TWa$ travelling wave is stabilised completely provided $\gamma\neq 0.$ The right panel shows the convergence of the phase speed $c$ using the descent method of equation (\ref{eq:s}). When $Re=200$ the travelling wave is no longer stabilised.
}
\end{centering}
\end{figure}

We compute $TWa$ using the TDF stabilisation method, the results shown in figure \ref{fig:TW} at $Re=40\, \& \, 100$ using $G_{max}=100,$ $\gamma=0.05,\, s(0)=0,$ $\kappa=1$ and $T=0.1.$ {Note, this case again sets $T_{start}=50$ giving us a turbulent initial condition before TDF is turned on.} These parameter values were arrived at after a very short amount of trial and error; in fact this travelling wave is stabilised over a large range of parameters at these Reynolds numbers. At $Re=40$ we demonstrate the effect of removing the adaptive shift by setting $\gamma=0,$ this also shows that the laminar solution is not stabilised in agreement with the results of the previous section. We notice that the dynamics are steady with the kinetic energy $E$ settling onto a value close, but not equal, to that of $TWa$ (the same is true for $D$ and $I,$ not shown for brevity) and the size of $Q$ tends to a small non-zero value. Our interpretation here is that $TWa$ is partially stabilised (because $c$ is small in this case) but $Q$ cannot tend to zero as $s$ is incorrect, thereby leaving some invasive energetics in effect. By setting $\gamma=0.05,$  $s$ adaptively adjusts such that $c=0.00198$ which is in agreement with the value reported in \cite{Chandler}, at the same time $Q$ drops to machine precision.

The result is repeatable at $Re=100$ with the rest of the parameters held fixed, only now the convergence rate is decreased. At $Re=200$ stabilisation is not found, even on increasing to $G_{max}=5000$. To understand the issue here we converge the $TWa$ solution at $Re=100$ using NGh (using the code from \cite{Chandler} and \cite{Lucas:2015gt}) and perform arc-length continuation in $Re.$ At $Re=100$ and $Re=200$ we then conduct a stability analysis of the solution via Arnoldi iteration, the unstable part of the spectrum is shown in figure \ref{fig:TW_eig}. The important feature to note is that the unstable travelling wave has gained an unstable direction at $Re=200$ with a purely real eigenvalue, thus violating the so-called `odd-number limitation'. This is why the stabilisation has been unsuccessful.

{Not to be deterred, we should consider possible ways to avoid this issue. In the case of the laminar solution in section \ref{sec:lam} the odd-number issue was avoided by taking advantage of the symmetry of the solution and adding a symmetry operator, in that case a translation in $x,$ into the TDF terms. For $TWa$ the continuous $\mathcal{T}$ symmetry has been broken, however on close inspection the solution is, in fact,  invariant under a $\mathcal{R}\mathcal{S}^3\mathcal{T}_\pi$ symmetry operation, see figure \ref{fig:TW_eig} (right).} 

We can, therefore, attempt an adjustment of the feedback forcing for $TWa$ with this additional symmetry imposed, i.e.
\begin{align}
f &= G(t)(\psi(x,y,t) - \mathcal{R}\mathcal{S}^3\mathcal{T}_\pi\psi(x-s,y,t-T)) \nonumber \\
 &= G(t)(\psi(x,y,t) + \psi(x-\pi-s,-y-\frac{3\pi}{4},t-T)) \label{eq:f_sym}
\end{align}

\begin{figure}
\begin{centering}
 \includegraphics[width=0.48\linewidth]{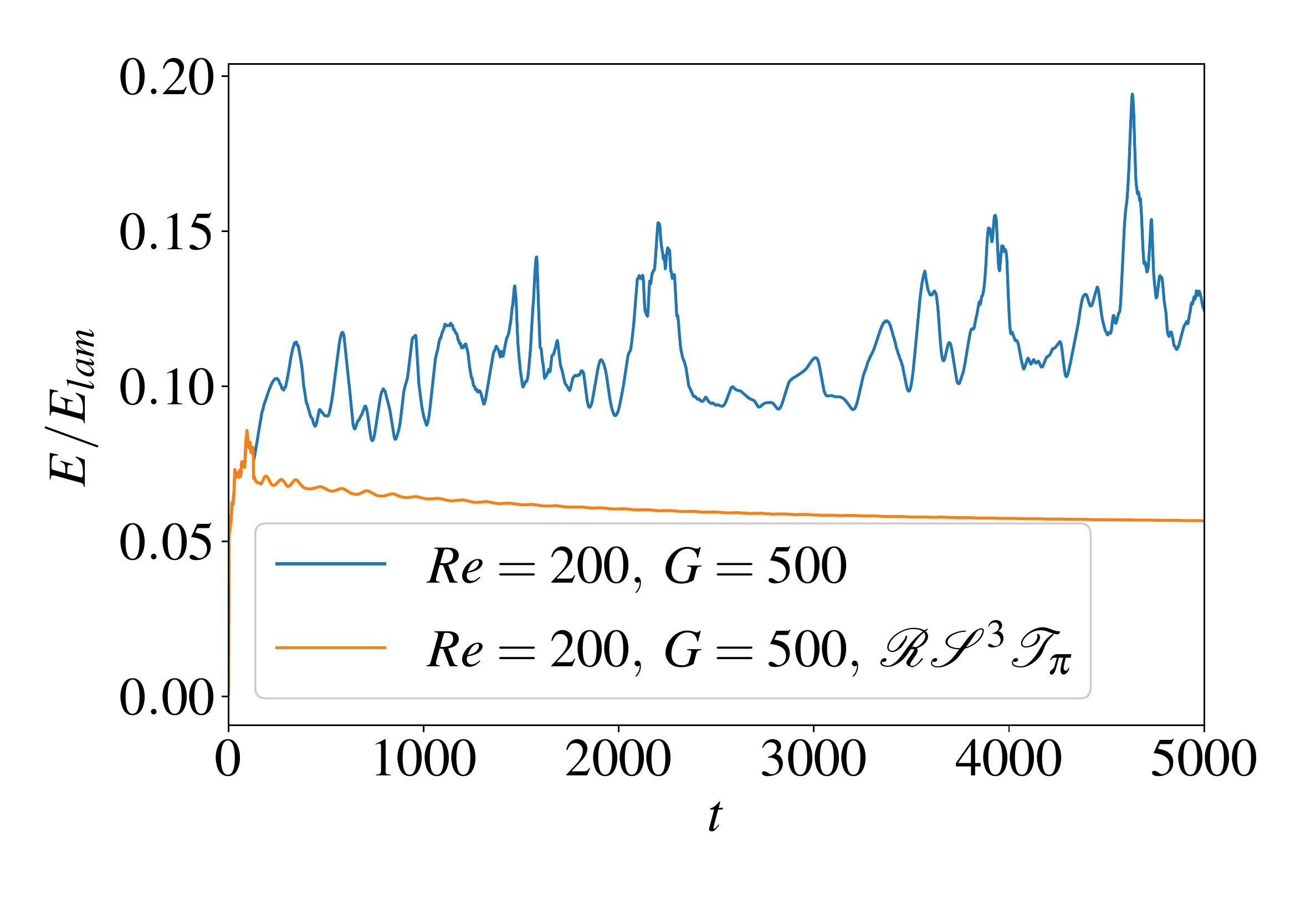}
 \includegraphics[width=0.48\linewidth]{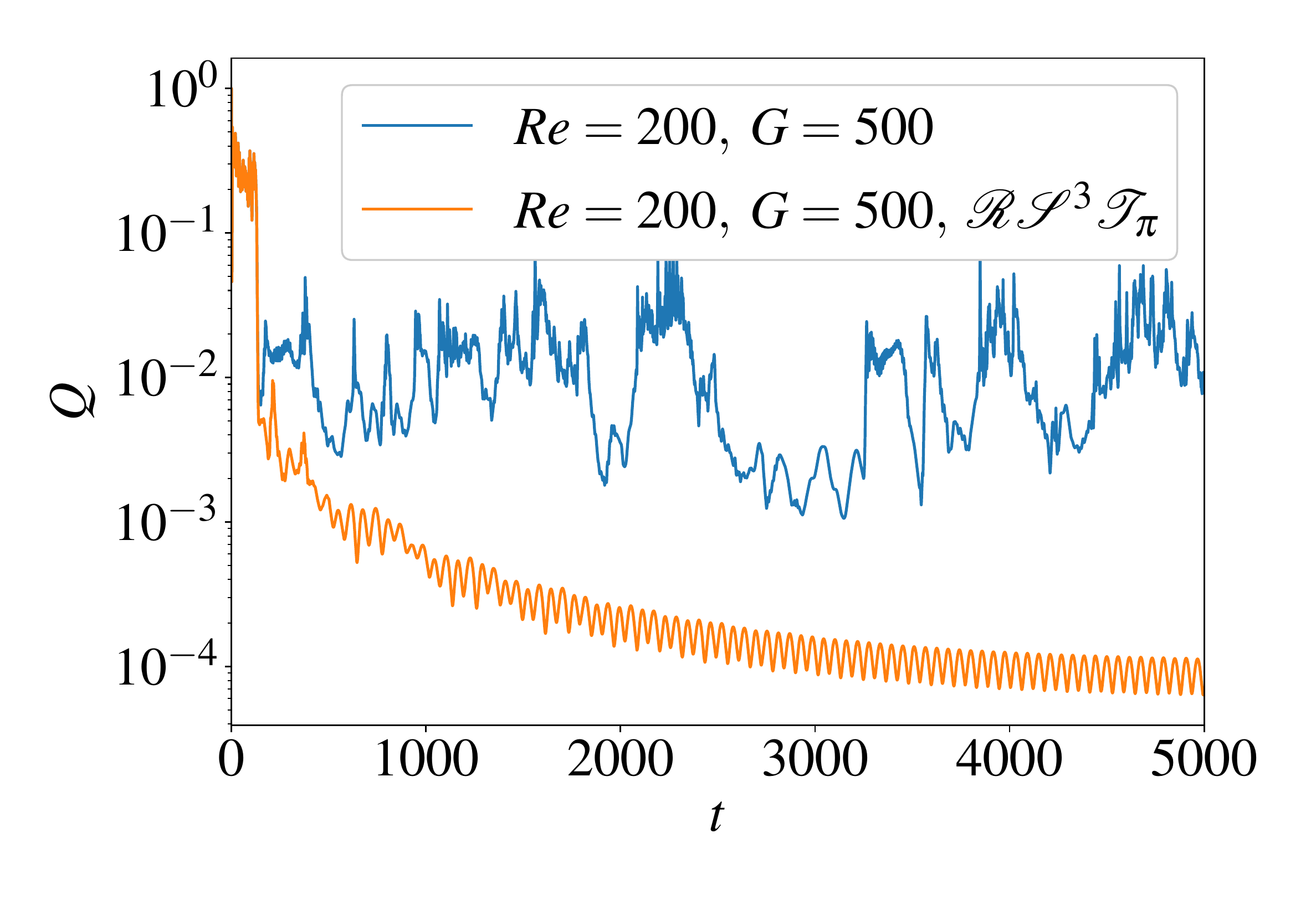}
 \caption{\label{fig:TW_200} TDF stabilisation of the $TWa$ travelling-wave solution of Kolmogorov flow shown via time series of $E/E_{lam},$ (left) and $Q$ (right) when applying TDF at $Re=200,\,\, G=500,\,\, \gamma=0.1$ in the case of applying the symmetry (orange) and not (blue). Stabilsation of $TWa$ is found to be successful, and the odd-number limitation is overcome, when applying the additional symmetry operation in the delay term, as in equation \eqref{eq:f_sym}. Note $Q$ includes the symmetry as in equation \eqref{eq:Qsym}.}
\end{centering}
\end{figure}

\begin{figure}
\begin{centering}
 \includegraphics[width=\linewidth]{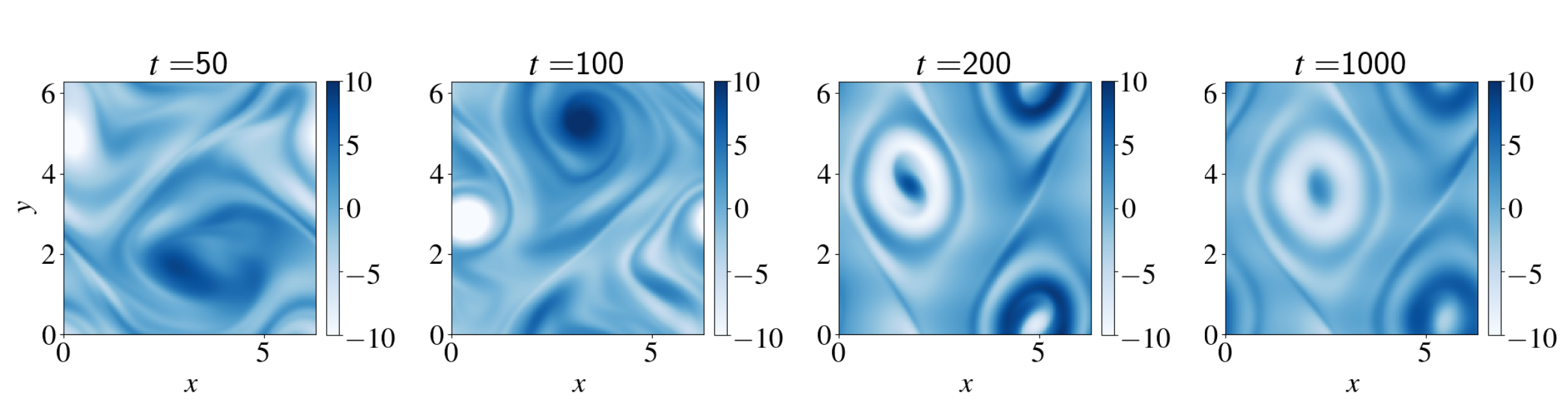}
\caption{\label{fig:vort_TW} Vorticity, $\omega,$  snapshots for the $Re=200$ stabilisation of the $TWa$ travelling wave with TDF using the symmetrised forcing as in equation \eqref{eq:f_sym}.  
}
\end{centering}
\end{figure}

We find that stabilisation is recovered at $Re=200,$ $G=500$ and $\gamma=0.1.$ Figure \ref{fig:TW_200} shows the energy $E$ and the residual $Q,$ now with the symmetry included, i.e.
 \begin{align}
Q(t) &= \frac{\left<\left(\psi - \mathcal{R}\mathcal{S}^3\mathcal{T}_\pi\psi(x-s,y,t-T)\right)^2\right>_V^{1\over 2}}{\left<\psi^2\right>^{1\over 2}_V}. \label{eq:Qsym}
\end{align}
Figure \ref{fig:vort_TW} shows vorticity snapshots for the stabilised case demonstrating the evolution from a highly disordered turbulent flow at early times to the more ordered but still nonlinear travelling wave solution (movie available in the supplementary material \cite{SupMovie}).

{Of course \emph{a priori} there was no guarantee such a change would have the required effect on the unstable spectrum, therefore we should consider how the application of this symmetry operator in TDF has recovered stabilisation of $TWa$. It is evident that the forcing terms \eqref{eq:f_sym} will drive the solution into, or at least towards, the $\mathcal{R}\mathcal{S}^3\mathcal{T}_\pi$ symmetric subspace. Inside that subspace the $TWa$ solution will have different symmetry properties. We can check the stability properties in the subspace by Arnoldi iteration while projecting the vorticity field, $\omega = \left(\omega + \mathcal{R}\mathcal{S}^3\mathcal{T}_\pi \omega \right)/2.$ The unstable eigenvalues are shown in figure \ref{fig:TW_eig}. While the travelling wave remains highly unstable in the symmetric subspace, the purely real eigenvalue is no longer present, explaining why TDF is successful. 
An obvious alternative having observed how the symmetry avoids the odd-number limitation, is to conduct TDF DNS in the projected subspace using the original TDF forcing \eqref{eq:TDF_f}. Results, not shown, confirm the solution is indeed stabilised with such a projection approach. However this arrangement is not in-keeping with the spirit of TDF where we are, in general, hoping to stabilise solutions which may be time periodic, satisfying some recurrence condition
$$ \omega(x,y,t) - \mathcal{R}^i\mathcal{S}^m\mathcal{T}_s\omega(x,y,t-T) = 0.$$
Notice this condition does not necessarily restrict the dynamics into the symmetric subspace, but certainly will for a steady solution. 
We also note that, similarly to the description given in section \ref{sec:lam} when analysing the stability of the laminar solution, the form of equation \eqref{eq:f_sym} is important. Applying the symmetry operation to the first term results in destabilsation and reversing the sign of $G$ results in the `critical level' instability. Here the effect is not as simple to interpret as the laminar solution since the symmetry operation, the underlying solution and its unstable manifold are all far more complicated. However by a simple analysis of equation \eqref{eq:f_sym} it can be seen that applying the symmetry operation to the term at $t$ rather than at $t-T$ will result in a perturbation away from a symmetric solution only being `removed' from the flow if the perturbation itself is also symmetric; of course not the generic situation.
} 

This naturally opens up the possibility of stabilising other travelling wave, or equilibrium solutions by using their symmetries. Moreover using the symmetries of the solution offers a simple way to constrain TDF to avoid previously stabilised solutions. It should be noted that the other known steady solutions in this system all suffer from the odd-number limitation at these Reynolds numbers, having at least one unstable eigenvalue on the real axis.  The work of Farazmand \cite{Farazmand:2016hf} contains possibly the most comprehensive study of steady and travelling wave solutions in this flow, making joint use of adjoint-descent and NGh to find highly unstable and rarely visited, high dissipation equilibria and travelling waves (16 equilibria and 9 travelling waves, including $\omega_{lam}$ and $TWa$). The majority of those reported (i.e. tables 1 and 2 in \cite{Farazmand:2016hf}) have a largest unstable direction which does not oscillate (has a purely real eigenvalue). Fortunately the majority of these solutions also lie in a symmetric subspace; this suggests that the odd-number limitation might be avoided by imposing a symmetry in the delay term of TDF. 


\section{Symmetries in TDF}\label{sec:sym1}

\subsection{Systematic search}\label{sec:sym}
We now present the results of a systematic investigation of the various choices of symmetry which can be imposed in equation \eqref{eq:f}, while keeping the delay $T$ small. Rotation $\mathcal{R}$ forms a cyclic group of order 2 and shift-reflect $\mathcal{S}$ forms a cyclic group of order $2n,$ meaning there are $4n$ distinct discrete symmetries. It should be noted that $\mathcal{R}\mathcal{S} \ne \mathcal{S}\mathcal{R}$ however $\mathcal{S}^{2n-1}\mathcal{R} = \mathcal{R}\mathcal{S}$ and $\mathcal{S}\mathcal{R}\mathcal{S} = \mathcal{R}.$ Therefore to consider these $4n$ discrete symmetries and the continuous symmetry $\mathcal{T}$ we will set the TDF force as
\beq
f = G(t)\left(\psi(\bm{x},t)-\mathcal{R}^j\mathcal{S}^m\mathcal{T}_s\psi(\bm{x},t-T)\right),
\eeq
with $j=0,1$ and $m=0,1,2n-1$ giving the $4n$ discrete symmetries and $\mathcal{T}_s$ giving the continuous symmetry. To avoid excessive calculations and noting that the adaptive method to converge $s$ described previously should self-select the required translations, we will only consider only five starting translations $s = 2\pi/2^n, n=0,1,2,3,4$. For this set of calculations we modify the form of the gain, now using

\beq G(t) = \min(G_{max},\kappa(t-T_{start})^2), \label{eq:G2}\eeq
instead of \eqref{eq:G}. This has a slightly smoother profile, enabling $G_{max}$ to be reached more quickly without undesirable long-lived invasive behaviour. It should be said that this is a minor improvement and many of the results presented would be reproduced using \eqref{eq:G}. {A hyperbolic tangent or sigmoid function would arguably be smoother still but as this quadratic works satisfactorily we leave a trial of other profiles for future work.}

\begin{figure}[!t]
 \centering
 \begin{minipage}{0.3\linewidth}
 \centerline{(a)}
 \includegraphics[width=\linewidth]{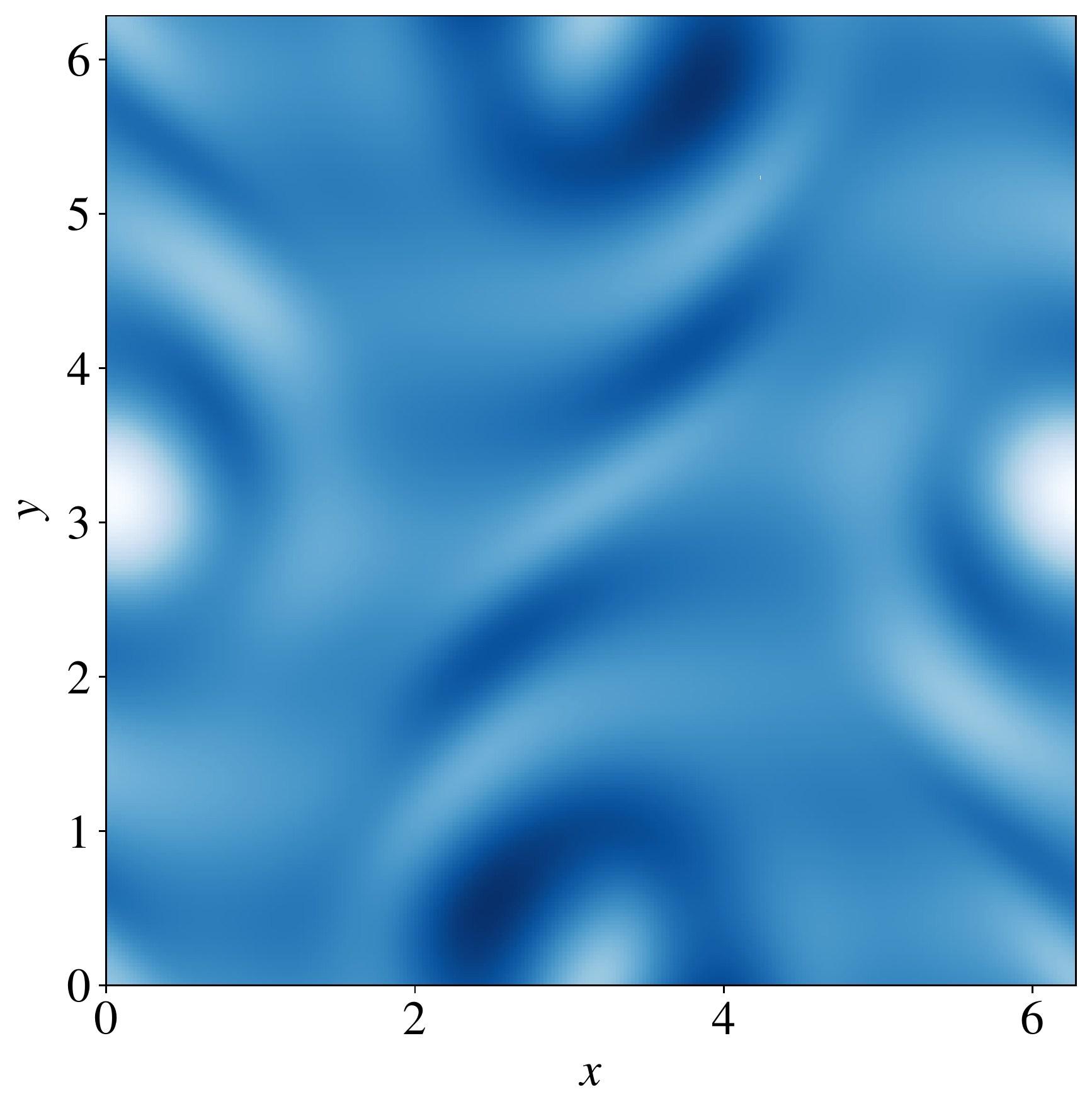}
 \end{minipage}
 \begin{minipage}{0.3\linewidth}
 \centerline{(b)}
 \includegraphics[width=\linewidth]{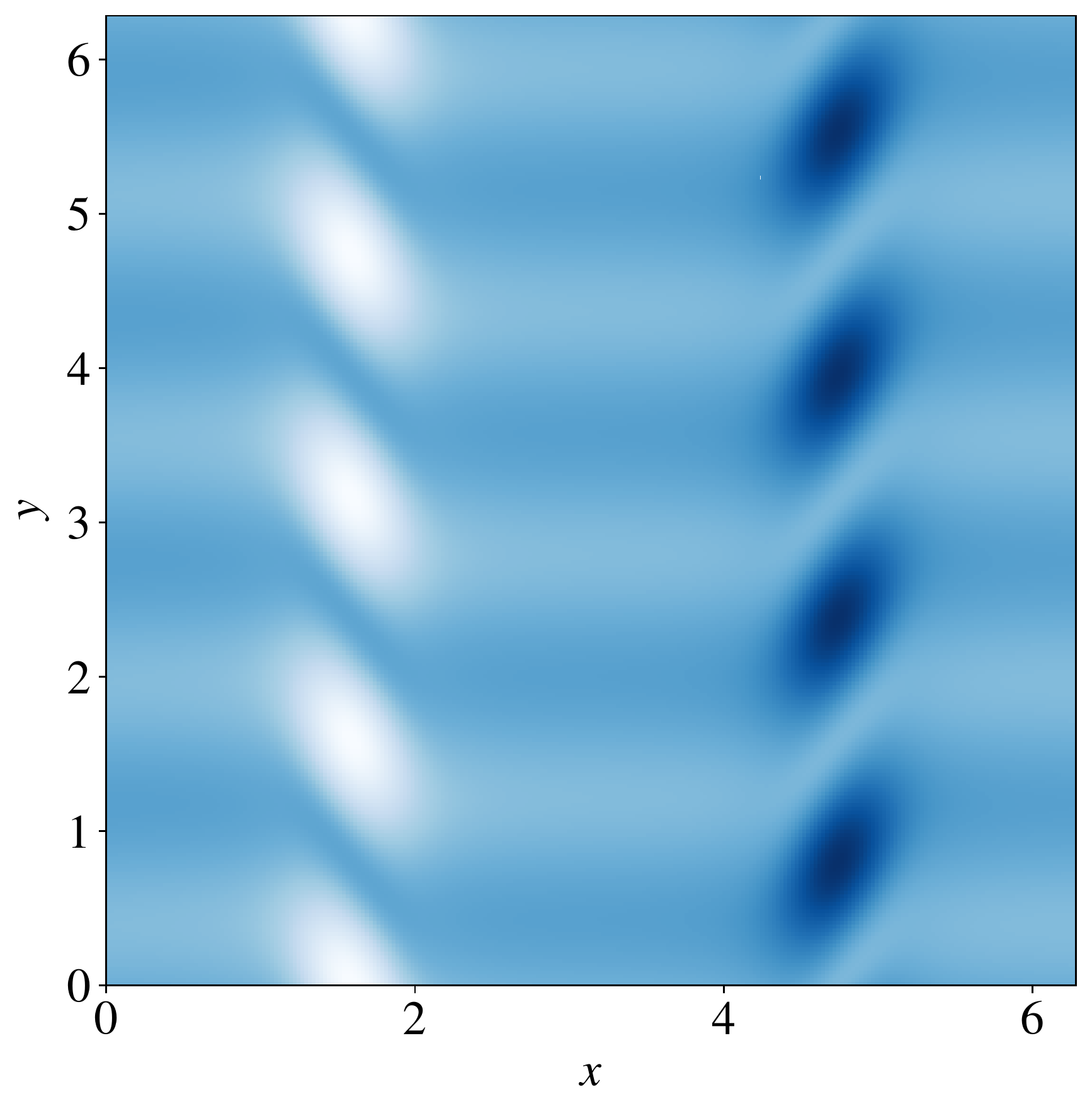}
 \end{minipage}
 \begin{minipage}{0.3\linewidth}
 \centerline{(c)}
 \includegraphics[width=\linewidth]{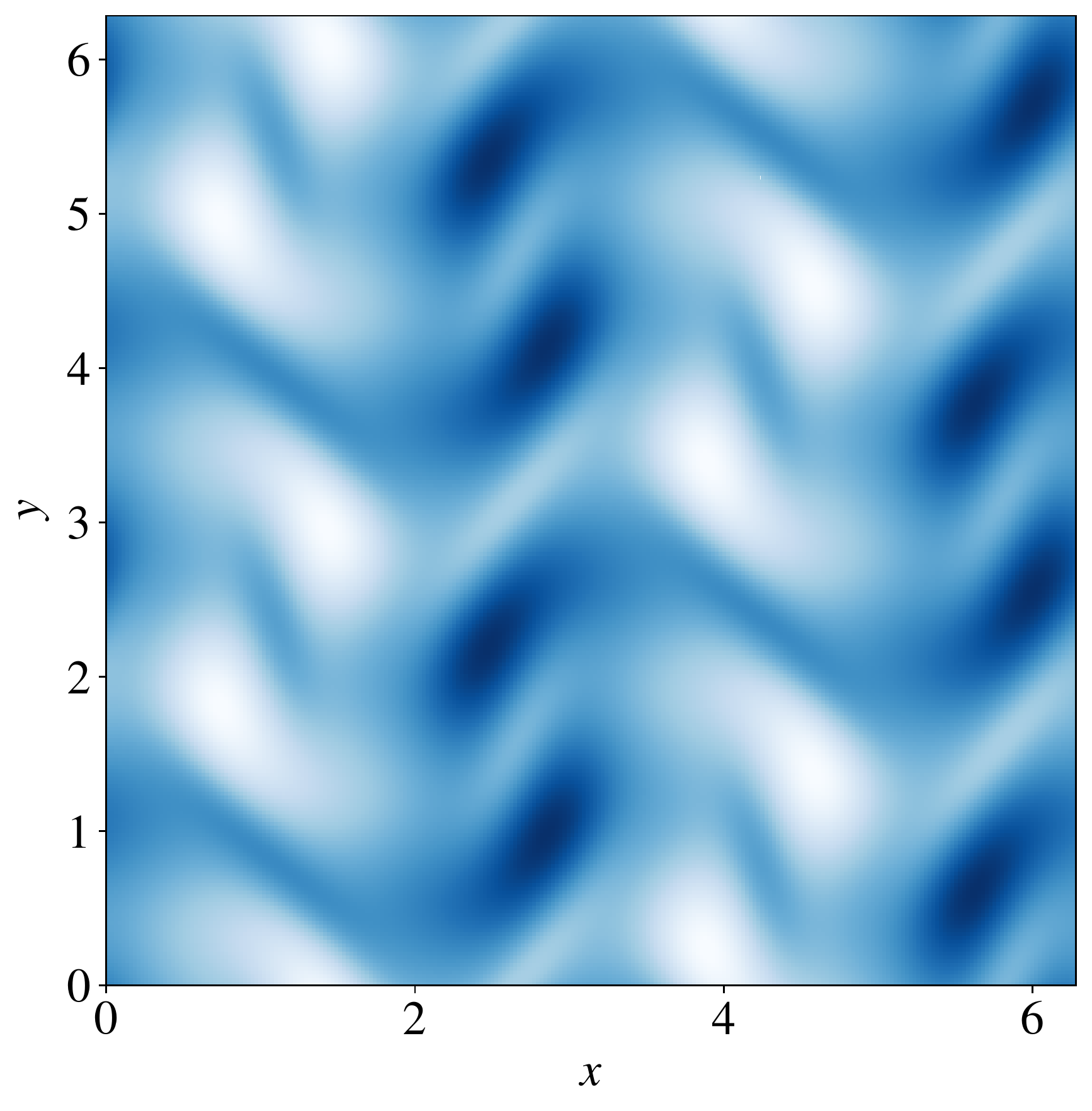}
 \end{minipage}
 \begin{minipage}{0.3\linewidth}
 \centerline{(d)}
 \includegraphics[width=\linewidth]{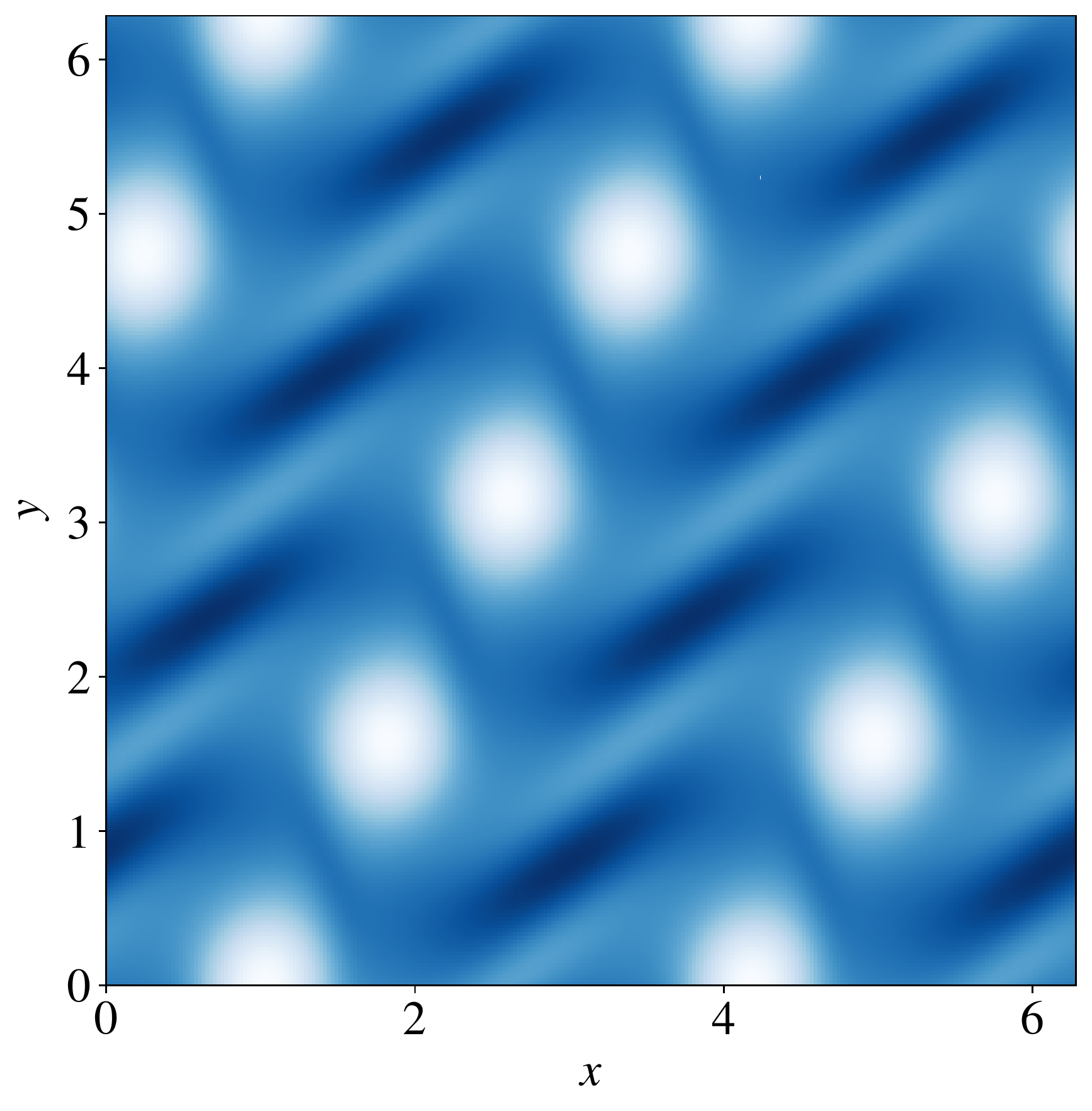}
 \end{minipage}
 \begin{minipage}{0.3\linewidth}
 \centerline{(e)}
 \includegraphics[width=\linewidth]{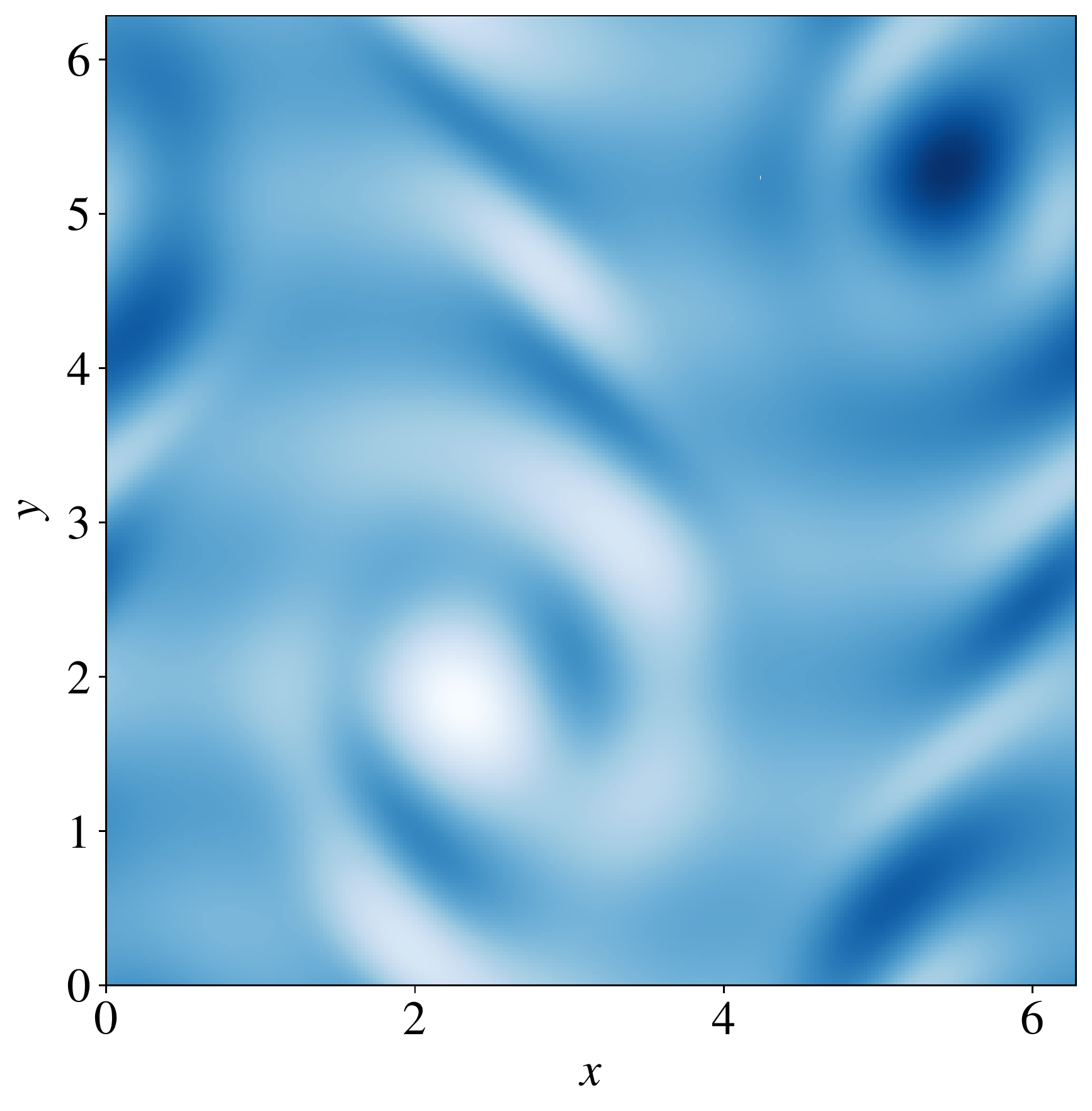}
 \end{minipage}
 \begin{minipage}{0.3\linewidth}
 \centerline{(f)}
 \includegraphics[width=\linewidth]{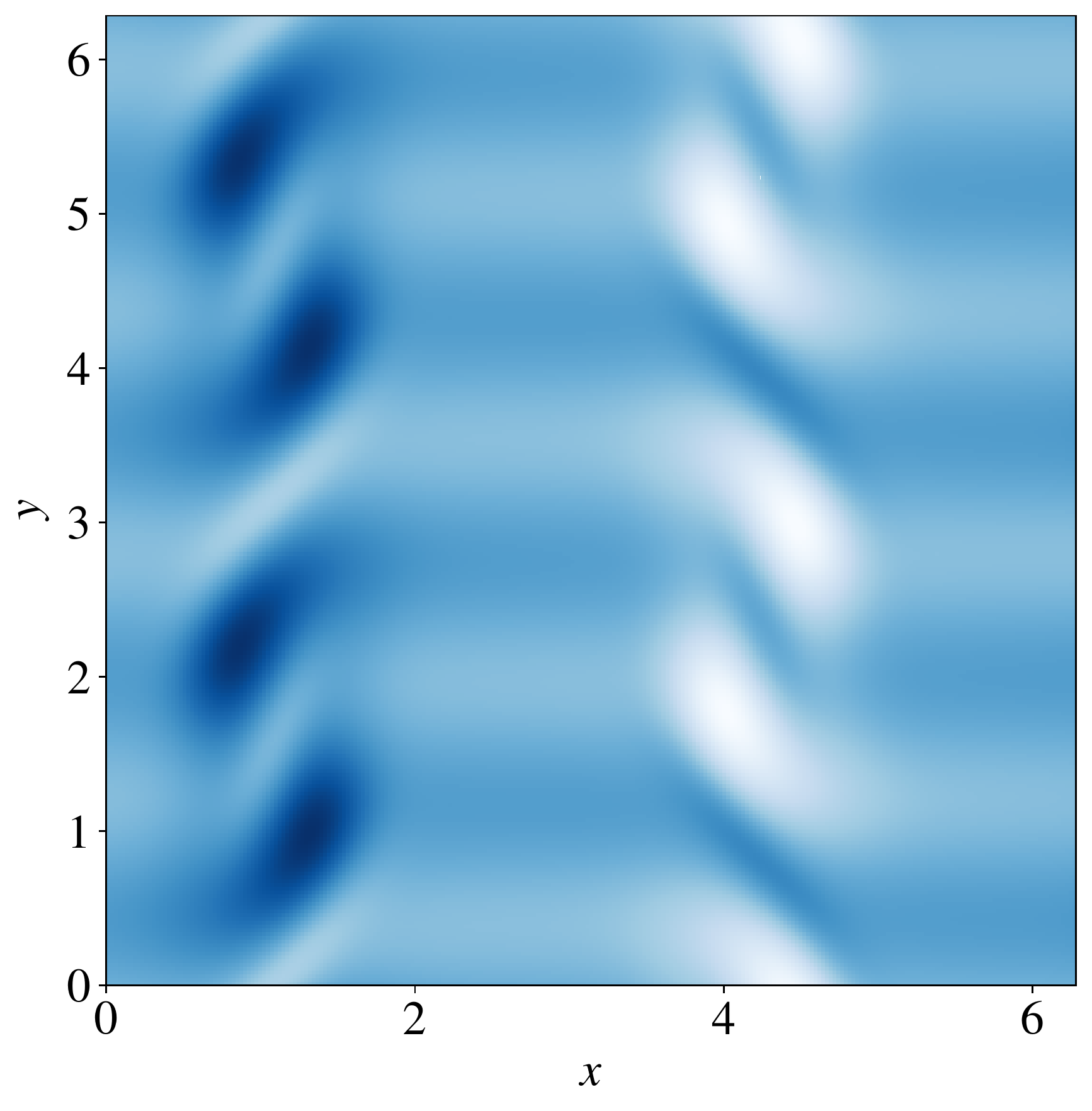}
 \end{minipage}
 \caption{
 Snapshots of  equilibrium solutions (a,b,c,d) and travelling-wave solutions (e,f) stabilised with TDF (see table \ref{tab:ECS}).  (a) EQa.   (b) EQb. (c) EQc. (d) EQd. (e) TWa. (f) TWb. \label{fig:vortECS}}
 \end{figure}

We will consider $Re=40,$ $G_{max}=20,$ $\kappa=0.2$ and $\gamma=0.05$. 
The result is 80 simulations in which 41 resulted in successful stabilisation, finding four equilibria and two travelling waves, including $TWa$ as described in the previous section, and summarised in table \ref{tab:ECS}. In all cases the solution is confirmed via convergence with NGh in the absence of TDF. Surprisingly only two of these other solutions have previously been reported in the literature; $EQa$ being the $E_1$ solution reported in \cite{Farazmand:2016hf}, and $EQb$ is the solution emanating from the primary bifurcation, discussed in \cite{Page2021} and equivalent to the $\alpha=1$ kink-antikink solution reported in \cite{Lucas:2014ew} in large domains. We discover quite a variety of flow structures in these stabilised ECSs as shown in the vorticity plots in figure \ref{fig:vortECS}. 


There is also some slight subtlety regarding the effect that continuous translations have on the TDF outcomes. In the event of stabilising a travelling wave, the translation must satisfy the $s=cT$ condition, however here are two further effects at play in general which depend on the flow structures and other symmetries involved. The first is, for equilibria, a pinning of the flow structures relative to the axis of reflection/rotation when they appear in combination. For example $EQb$ is repeatedly stabilised for various translations when $\mathcal{S}^3$ or $\mathcal{S}^7$ is applied. These cases simply arrive at different ``copies'' of $EQb$ with the kink-antikink structure pinned at the $x$ position given by $s,$ respecting the shift-reflect (figures in the supplementary material \cite{SupData}). This is the reason why we see such a large number of repeated stabilisations of the same solution when changing $s$ in the presence of either an odd shift-reflect or rotation. 
Alternatively with no reflect or rotation the translation can produce ECSs with repeated flow structures within the domain, for example $EQc$ has a clear mode 2 pattern due to the $s=\pi$ translation imposed, likewise $EQd$ has an unusual diagonal array of vortices which respect the  $\mathcal{S}^2$ or $\mathcal{S}^6$ shift in $y$ (so no net reflection) and $s={\pi \over 4}$ in $x.$


We also need to take some care in monitoring the outcomes of applying TDF; of the 39 we classify as unsuccessful we find 27 of these tend to a steady state, but $Q$ does not vanish meaning TDF is invasive and the steady solution is not a solution of the Navier-Stokes equations. A secondary way to identify such cases is that $I\neq D$ meaning that the perturbing force is playing a role in the energetic balance. The other 12 cases are unsteady, either periodic, quasi-periodic or fully chaotic. More details can be found in the supplementary material \cite{SupData}. In figure \ref{fig:symT} we show the time-series plots of the energy $E(t),$ the `residual' $Q(t)$ and the relative error  $|I-D|/|I|$ for four cases; $\mathcal{T}_\pi$ where the evolution remains chaotic; $\mathcal{ST}_\pi$ where the flow becomes steady but the TDF term does not vanish and $I\ne D;$ 
$\mathcal{S}^4$ which stabilises $TWb$ and $\mathcal{S}^5$ which stabilises $EQb.$ It can be seen that the travelling wave stabilisation takes longer for $Q$ to vanish as it requires the additional condition that $s = cT$ to be satisfied through the solution of \eqref{eq:s}.

\begin{figure}[]
\centering
 \begin{minipage}{0.45\linewidth}
 \centerline{$\mathcal{S} \mathcal{T}_\pi$} \includegraphics[width=\linewidth]{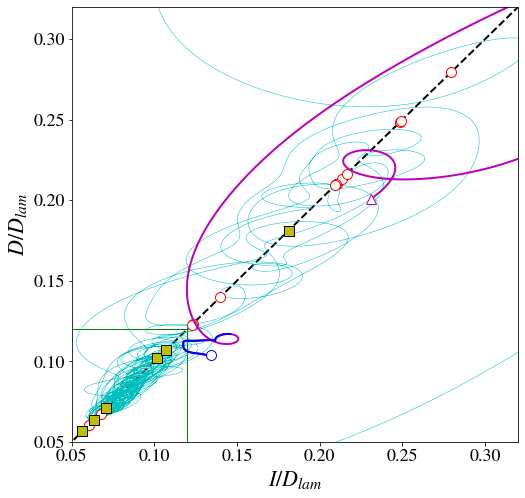}
\end{minipage}
 \begin{minipage}{0.45\linewidth}
 \centerline{$\mathcal{S}^5$} \includegraphics[width=\linewidth]{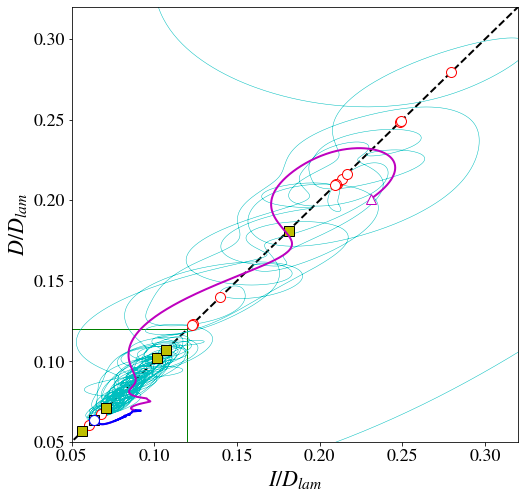}
 \end{minipage}

\caption{Plots of the $(D,I)$ projection for the TDF cases with symmetries $\mathcal{S} \mathcal{T}_\pi$ (left) showing the trajectory tending towards an `invasive' steady state where $D\neq I$ and $\mathcal{S}^5$ (right) showing successful stabilisation of $EQb.$ $t<T_{start}$ is shown in cyan, $T_{start}+10>t>T_{start}$ magenta and $t>T_{start} + 10$ blue. States from table \ref{tab:ECS} are shown as yellow squares and states in \cite{Farazmand:2016hf} as red circles. 
\label{fig:DI_ECS}
}
\end{figure}

Figure \ref{fig:DI_ECS} shows the projection of the dynamics on the plane $(I/D_{lam},D/D_{lam}),$ the left panel showing a failed stabilisation and invasive steady state when applying $\mathcal{S}\mathcal{T}_{\pi},$ right panel showing the stabilisation of $EQb$ when applying $\mathcal{S}^5.$ This figure demonstrates the quite different dynamics in each case, at least on this projection, and that the stabilised state can be quite far from the location in phase space at $T_{start}$ (magenta triangle in the figure) suggesting that the basin of attraction in this case is large. 

Table \ref{tab:ECS} also reports the stability of the ECS, in particular noting the largest unstable eigenvalue, the dimension of the unstable manifold and the number of those directions which do not oscillate. We find all solutions, except $TWa,$ have multiple directions violating the odd-number limitation, demonstrating that the use of symmetries has overcome this issue. {We have also confirmed that projection into the symmetric subspace will remove the real unstable eigenvalues for these solutions. Note that this means that $EQb$ and $TWb$ are stable in their respective subspaces, having only purely real unstable eigenvalues.}


\subsection{The limit $T\to 0$}\label{sec:T}
\begin{figure}[]
\centering
\includegraphics[width=\linewidth]{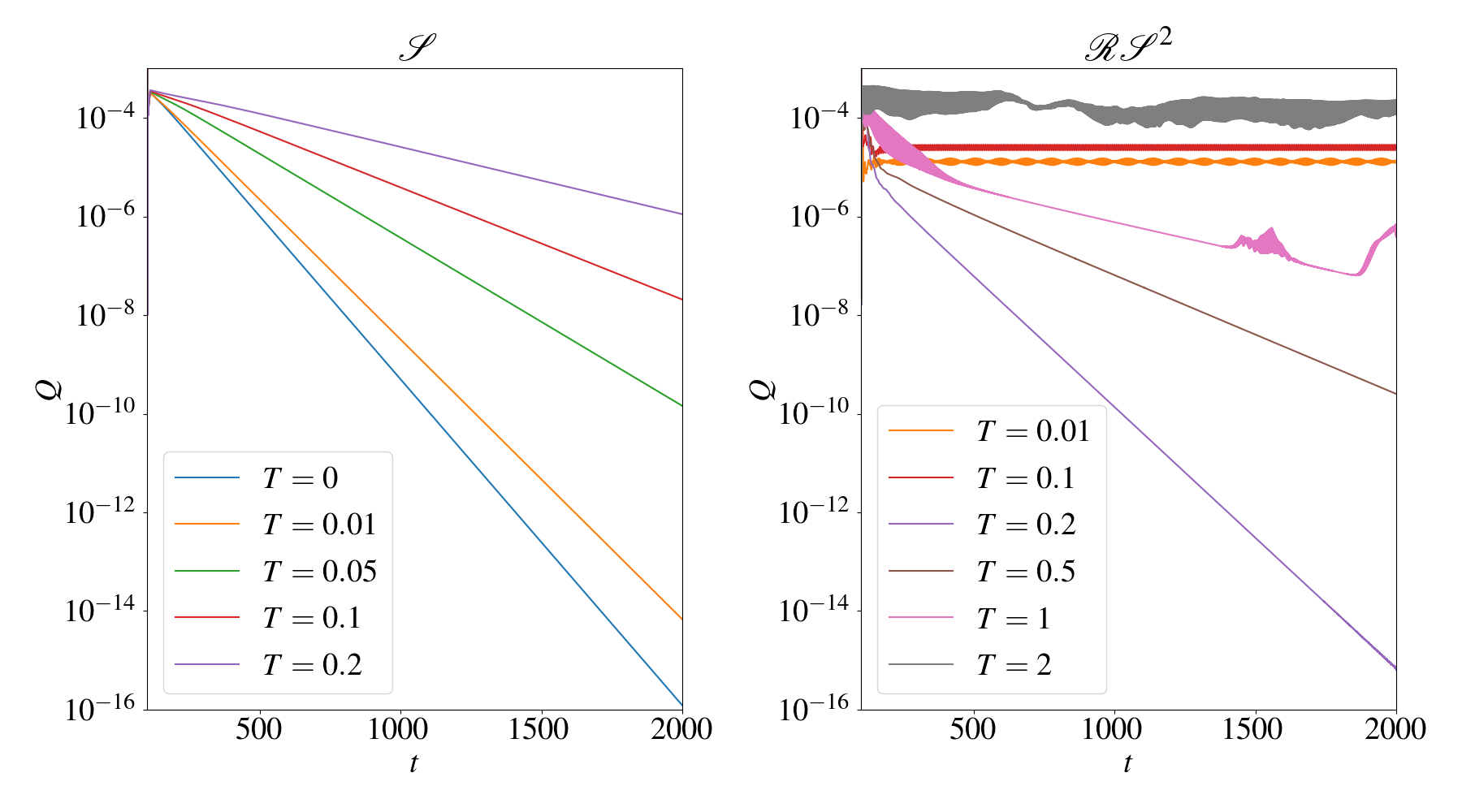}
\caption{
Plot of $Q(t)$ for TDF at $Re=40,\, G_{max}=20$ using the symmetry $\mathcal{S},$ stabilising the equilibrium $EQ_b$ (left) and $\mathcal{RS}^2$ stabilising $EQa$ (right) for different values of the time delay $T.$ We see that smaller delay periods result in faster attraction to the $EQa$ and that no delay $T=0$ is most effective. However when applying $\mathcal{RS}^2$ $EQa$ is only stabilised for intermediate values of $T.$ \label{fig:Q_T}
}
\end{figure}

In section \ref{sec:lam} it was noted that, when using additional symmetry operations embedded in the TDF term, the principle effect of TDF on the eigenvalue spectrum can persist as $T\to 0.$ While this limit is irrelevant for travelling wave solutions, as the $c=s/T$ combination must be satisfied in this formulation, it remains to be confirmed if the stabilisation of equilibria reported in table \ref{tab:ECS} remains in this limit. The left panel of figure \ref{fig:Q_T} shows the effect of varying $T$ when stabilising $EQb$ by including the $\mathcal{S}$ symmetry in TDF ($j=0,\,m=1,\, s=0$ in equation \eqref{eq:f_sym}). The rate of stabilisation increases as $T$ decreases; finite $T$ slows the stabilisation and the solution can be stabilised with no delay. This is consistent with the interpretation made in section \ref{sec:lam} that the contribution involving $T$ (for small $T$) will scale the spectrum toward the origin; when the whole spectrum is to the left of the imaginary axis this means that the least unstable mode becomes smaller in absolute terms (less negative), resulting in a slower decay as $T$ increases. 

However introducing the rotational symmetry and attempting to stabilise $EQa$ with $\mathcal{RS}^2$ (right panel figure \ref{fig:Q_T}) we find the stabilisation does not persist in this limit. In this case the rotation in physical space will introduce anti-diagonal entries in both the matrices $A$ and $B$ \eqref{eq:A}-\eqref{eq:B} making the interpretation of the TDF effect much less trivial. In this case we find that small $T$ is unable to stabilise the solution (with $G_{max}=20$) and likewise large $T$ introduces unsteady oscillations, therefore there is an optimal $T$ at these parameters of $T\approx 0.2.$ 

{The reason for this difference can be found when examining the eigenvalue spectrum for these solutions following projection into the respective symmetric subspaces. Note that $EQb$ has a purely real unstable spectrum. Following projection into the $\mathcal{S}$ subspace, this solution is stable. This means that actually TDF is doing nothing other than drive the solution into the subspace and the time-delay is redundant. On the other hand $EQa$ is not stable in the $\mathcal{RS}^2 $ subspace, having 4 complex unstable eigenvalues. This means the time-delay is required to complete the stabilisation of this solution.}

\begin{table}
\begin{tabular}{lcccccc p{4cm}}
 & $c$      & $I=D$    & $E$     & $\lambda_R$  & $\lambda_I$ & $N$ & Symmetry \\
\hline \hline
$EQa$  &  0       & 0.1273    & 0.7615  & 0.2181   & 1.137                 & 6  (2)   &  $\mathcal{R}\mathcal{T}_{*}, \, \mathcal{R}\mathcal{S}^2\mathcal{T}_{*}, \,\mathcal{R}\mathcal{S}^4\mathcal{T}_{*},\newline \mathcal{R}\mathcal{S}^5\mathcal{T}_{*},\mathcal{R}\mathcal{S}^6\mathcal{T}_{*}$\\
$EQb$  &  0       & 0.07953  & 0.6151  & 0.5961   & 0.0                   & 9  (9)  & $\mathcal{S}\mathcal{T}_{0,\pi/4},\, \mathcal{S}^2, \, \mathcal{S}^3\mathcal{T}_{*},\newline \mathcal{S}^6, \, \mathcal{S}^7\mathcal{T}_{*}, \,$ \\
$EQc$  &  0       & 0.1341   & 0.3927  & 0.3984   & 0.0                   & 9  (3)  &  $\mathcal{S}^2\mathcal{T}_{\pi},\,\mathcal{S}^6\mathcal{T}_{\pi}$\\
$EQd$  &  0       & 0.2265   & 0.5137  & 0.6235   & $ 0.0 $ & 13 (7) & $\mathcal{S}^2\mathcal{T}_{\pi/4},\,\mathcal{S}^6\mathcal{T}_{\pi/4}$  \\
\hline
$TWa$  &  0.01976 & 0.08861  & 0.6975  & 0.06828  & 0.3545                & 4  (0)  & 0  \\
$TWb$  &  0.00092 & 0.07059  & 0.5482  & 0.5002   & 0.0                   & 2  (2)  & $\mathcal{S}^4$  \\
\hline \hline
\end{tabular}
\caption{
List of equilibrium solutions (EQ) and travelling-wave solutions (TW) stabilised with TDF. The symmetries which stabilise the solution are given in the table; where multiple translations stabilise they are shown with either multiple subscripts on $\mathcal{T}$ or if many starting $s$ stabilise we denote with *. Full details available in the supplementary material \cite{SupData}. All solutions have been confirmed by convergence with NGh and their unstable directions computed via Arnoldi iteration. The leading unstable eigenvalue of the solutions is $\lambda = \lambda_R+\mathrm{i}\lambda_I$. Here $N$ is the number of unstable eigenvalues (dimension of the unstable manifold) and the value in round brackets denotes the number of purely real unstable eigenvalues.
}\label{tab:ECS}
\end{table}
\begin{figure}[]
\centering
\includegraphics[width=\linewidth]{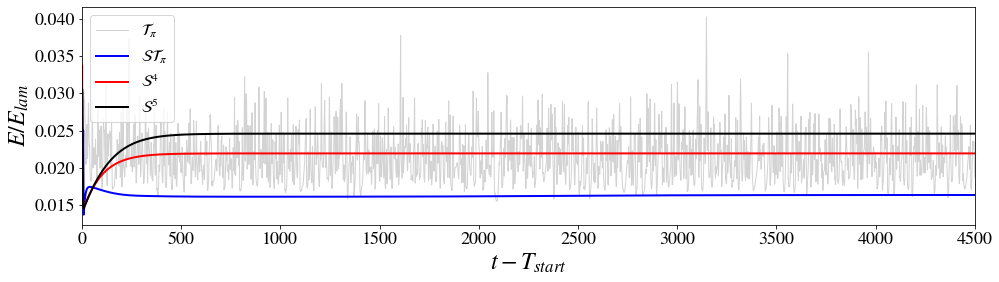}
\includegraphics[width=\linewidth]{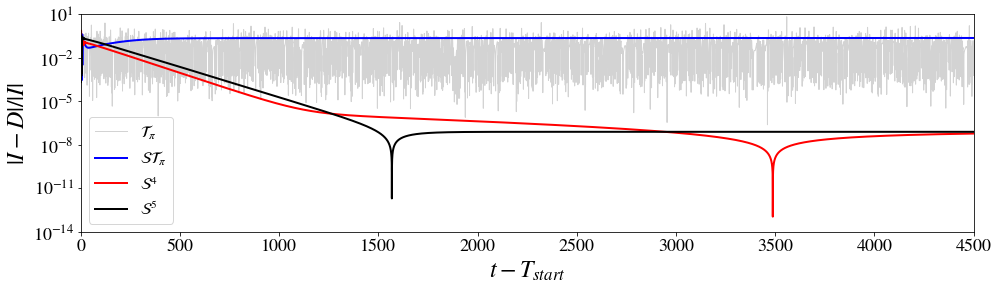}
\includegraphics[width=\linewidth]{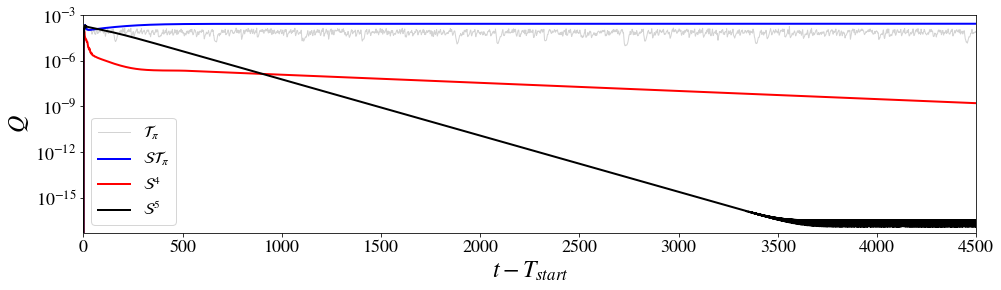}
\caption{
Plots of time series of
four simulations with different symmetries applied:
$\mathcal{T}_{\pi}$ (gray line, chaotic dynamics),
$\mathcal{S}\mathcal{T}_{\pi}$ (blue, steady state invasive TDF),
$\mathcal{S}^4$ (red, stabilisation of TWb),
$\mathcal{S}^5$ (black, stabilisation of EQb) .
Top $E(t)/E_{lam}$, middle $|I(t)-D(t)|/|I(t)|,$ bottom $Q(t)$.
Here, $T_{start}$ is the starting time of time-delayed feedback control.\label{fig:symT}
}
\end{figure}

\subsection{Multiple delays}
With the successes reported above we are  motivated to consider further generalisations of the method in order to stabilise more solutions and, moreover, avoid repeatedly stabilising the same ECSs reported in table \ref{tab:ECS}. An obvious generalisation for steady or travelling wave solutions with symmetries is to include more than one delay term with differing symmetry properties. In its most general form this would read


$$ f = \sum_i G_i(t)\left(\psi(\bm{x},t)-\mathcal{R}^{j_i}\mathcal{S}^{m_i}\mathcal{T}_{s_i}\psi(\bm{x},t-T_i)\right),$$

with different combinations of $j_i,\, m_i,\, s_i$ and the possibility for different delays $T_i$ and gains $G_i$ on each term. Clearly we can anticipate diminishing returns as more terms are included in this sum; solutions with more symmetries will typically be lower amplitude, in the sense of being closer to $\omega_{lam},$ and less likely to be embedded in the chaotic set. However given the symmetries displayed by solutions reported in the literature \cite{Farazmand:2016hf,Chandler}, and the number of repeated stabilisations in table \ref{tab:ECS}, it is quite common for steady solutions to retain more than one distinct symmetry. Adding these additional terms can permit further modification of the unstable manifolds, or symmetry constraints of the solutions albeit at the expense of extending an already high dimensional parameter space even further. As such we only show a preliminary test with two such terms here. 

Motivated by the figures shown in \cite{Farazmand:2016hf} where many of the ECSs have an $\mathcal{S}^4$ symmetry, we will consider a set of three further calculations with two TDF terms such that $\max(G_1) = \max(G_2) = 10$ and $T_1=T_2=0.2$ to be comparable with section \ref{sec:sym} (using \eqref{eq:G2} and $\kappa=0.2$). All cases will have one term with $j_1=0,\, s_1=0,\, m_1=4,$ the second term varying with the first case having $j_2 = 1,\, s_2=0,\,m_2=0,$ the second case $j_2 = 0,\, s_2=\pi, \,m_2=0,$ and the third case $j_2 = 1,\, s_2=\pi, \,m_2=0.$ In other words the first TDF term has $\mathcal{S}^4$ and the second term considers combinations of translation, $\mathcal{T}_\pi,$ and rotation, $\mathcal{R},$ leaving the shift-reflect zero. { Note in the cases with $\mathcal{T}_\pi$ we include this operator in addition to the dynamically adjusted translation described in section \ref{sec:TW}, which is applied to both terms in order to allow for travelling waves within the invariant subspaces.}

\begin{figure}[!t]
\centering
\includegraphics[width=0.9\linewidth]{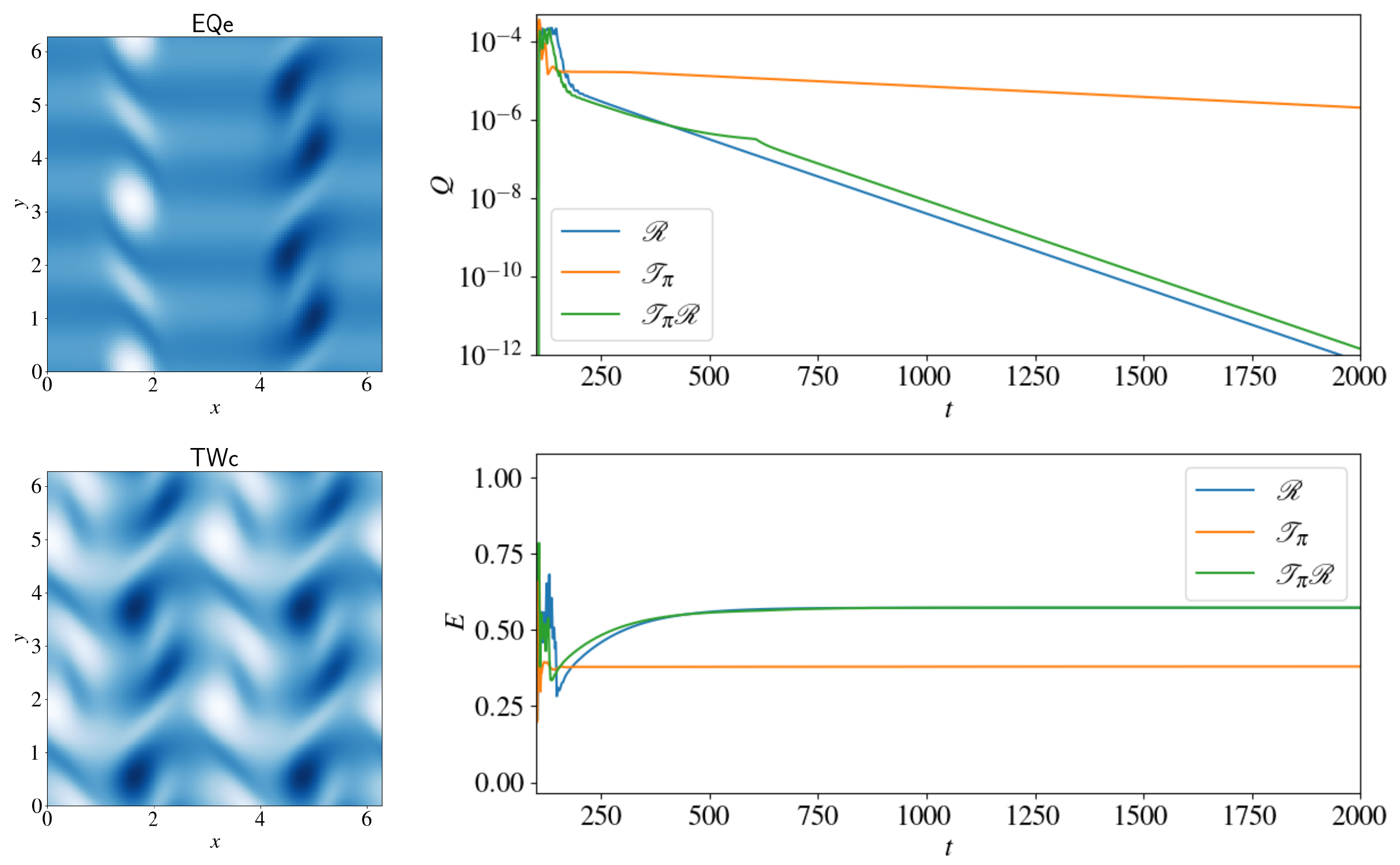}
\caption{
Plots of the further ECSs obtained using two terms of TDF as described in the text and outlined in table \ref{tab:ECS2}. Left shows the vorticity for $EQe$ (top) and $TWc$ (bottom) with, right, the time series of $Q(t)$ (top) and $E(t)$ (bottom) for the three symmetries on the second TDF term, $\mathcal{R}$ and $\mathcal{R} \mathcal{T}_\pi$ stabilising $EQe$ and $\mathcal{T}_\pi$ stabilising $TWc.$ \label{fig:vortECS2}
}
\end{figure}

This results in two further ECSs being stabilised, one equlibrium and one travelling wave, the equilibrium being E4 and the travelling wave being T3 from \cite{Farazmand:2016hf}. These solutions are summarised in table \ref{tab:ECS2} and figure \ref{fig:vortECS2}. Remarkably these solutions are different to either solution ($TWb$ or $EQa$) found when applying $\mathcal{S}^4$ \emph{together} with rotation and/or translation, demonstrating that there is scope for obtaining a range of solutions with similar or the same symmetries. 

{Using projection into the relevant combined symmetric subspace gives some additional interesting insight. Solution $TWc$ has its real eigenvalues filtered out by projection into the $\mathcal{S}^4$ subspace alone; the $\mathcal{T}_\pi$ subspace does not dramatically alter the solution's stability. This demonstrates that $TWc$ is, in fact, an attractor in the TDF  $\mathcal{S}^4$ system along with $TWb.$ We conjecture that $TWb$ has the larger basin of attraction and that the $\mathcal{T}_\pi$ symmetry is acting as a constraint to enable us to find $TWc$ in this case. It should be possible to find $TWc$ without the $\mathcal{T}_\pi$ TDF term but with a suitable initial condition using $\mathcal{S}^4$ alone. This observation opens up a number of interesting questions about dealing with multiple attractors when using TDF.}
\begin{table}
\begin{tabular}{lccccccc}
 & $c$      & $I=D$    & $E$     & $\lambda_R$  & $\lambda_I$ & $N$ & Symmetry 2\\
\hline \hline
$EQe$  &  0       & 0.0843    & 0.573 &  0.595  & 0    & 4  (4)   &  $\mathcal{R}\mathcal{T}_{\pi}, \, \mathcal{R} $\\
$TWc$  &  0.0183  & 0.1344  & 0.380 & 0.493   &  0  & 10 (3)  & $\mathcal{T}_{\pi}$ \\
\hline \hline
\end{tabular}
\caption{
Additional solutions stabilised with two TDF terms, one with $\mathcal{S}^4$ and the other with symmetry 2 shown in the table. All solutions have been confirmed by convergence with NGh and their unstable directions computed via Arnoldi iteration. The leading unstable eigenvalue of the solutions is $\lambda = \lambda_R+\mathrm{i}\lambda_I$. Here $N$ is the number of unstable eigenvalues (dimension of the unstable manifold) and the value in round brackets denotes the number of purely real unstable eigenvalues.
}\label{tab:ECS2}
\end{table}
\section{Summary \& Discussion}\label{sec:disc}

This study has shown several useful and original results applying time-delayed feedback control in two-dimensional turbulence. First is stabilisation of the laminar solution. 
  Despite the laminar state violating the so-called  `odd-number' limitation, we have taken advantage of the continuous symmetry of the solution to manipulate the linear operator and find stabilisation for certain choices of translation $s$ and gain $G.$ The DNS shows good agreement with the linear analysis.
We have also shown that by applying TDF in conjunction with an adaptive method for the translation $s,$ we are able to completely stabilise the $TWa$ travelling wave solution. This breaks down at high $Re$ where the solution gains a purely real unstable eigenvalue, however by once again using the symmetries of the solution we can avoid this `odd-number' limitation and stabilise the travelling wave up to $Re=200.$ {We discover that on projection into the symmetric subspace where this solution resides, the offending real eigenvalue is filtered out thereby enabling TDF to succeed again.} It should be emphasised at this Reynolds number the travelling wave has 36 unstable eigenvalues and we obtain it simply by timestepping the equations. 

Having found that TDF, {augmented with symmetries,} can successfully manipulate the unstable spectrum of ECSs, a systematic effort applying TDF with various symmetry combinations yielded 5 additional solutions at a single set of parameter values ($Re=40,\, G_{max}=20,\, \gamma=0.05$). While the success rate for the method, when trialling all possible symmetries, was not high ($\sim 50\%$), it should be said that we performed a systematic study to show the effect of applying all $4n$ discrete symmetries and 5 possible (starting) translations. In practice one would be unlikely to attempt this, recognising that typical flow structures, or alternatively the chaotic set, will be some distance from certain symmetric subspaces. For example, the inverse cascade of two-dimensional turbulence creates a large scale coherent structure in the form of a vortex dipole; such a structure is unable to be invariant under, say $\mathcal{R}\mathcal{S}$ or $\mathcal{R}\mathcal{S}^3$ (likewise for the kink-antikink structures exhibited in $EQb$). More selective choice of the symmetries to target relevant flow structures would improve the efficiency. 

{By computing the unstable directions of these solutions in the pertinent symmetric subspaces we find that the symmetry constraints serve to filter out the purely real unstable eigenvalues in the successful cases. Where the solutions have only real unstable eigenvalues this indicates that TDF is effectively constraining the dynamics into the symmetric subspace and this is sufficient for stabilisation. This is corroborated by the observation that, in such cases, the time-delay is not necessary for stabilisation. We therefore suggest that in future investigations where symmetric solutions are sought from scratch (i.e. not knowing their existence in advance) TDF offers a more general approach than projection alone. TDF will give you any solutions stable in the symmetric subspace, but it will also give you unstable solutions with complex eigenvalues (in the subspace). We hope that other methods known to avoid the odd-number limitation may offer further improvements \citep{Schuster1997,Pyragas:2006gv,Flunkert:2011cg}.}


A demonstration of generalising this approach by adding additional TDF terms showed that two more solutions can be stabilised at similar parameter values when trialling only three more carefully chosen symmetry combinations. This generalised TDF is similar in spirit to the extended TDF of \cite{Socolar1994, Pyragas:2001ch} and may prove to be useful at higher Reynolds numbers where dimensions increase and a broader range of spatiotemporal scales are active. {We note that one finding of these results was that with TDF and the $\mathcal{S}^4$ symmetry both $TWb$ and $TWc$ are stabilised. $TWc$ also has a $\mathcal{T}_\pi$ symmetry which does not help with the stabilisation in TDF but does constrain the flow structures to avoid re-stabilising $TWb.$ This shows another utility of using symmetries with TDF, in the future we should consider alternative ways of dealing with multiple attractors when stabilising solutions with TDF; it is clear from known solutions in the literature \citep{Farazmand:2016hf} that many ECSs lie in the $\mathcal{S}^4$ subspace.}

%




It should be emphasised that the success of stabilising solutions using their symmetries constitutes a work-around of the `odd-number' limitation discussed in the introduction. This is an important observation as it is possible the method has been neglected in the fluid mechanics community due to a presumption that this issue would be too restrictive.  The effect we find here has some similarity to other examples of avoiding this issue where a complex gain is used to modify the phase of the delay terms \citep{Fiedler:2007}. Having shown one resolution to this issue there is now reasonable motivation for attempting the method in other flows with different symmetry and bifurcation properties.  

There are numerous other avenues for future work on this method. For example we have paid little attention to the choice of $G(x,y,t);$ it is clear that too small a value will fail to stabilise, but also too large a value can cause invasive behaviour for long times, even if there is an ECS to be stabilised. Our results are the result of a small amount of trial and error to set a $G_{max}$ which was practical in this case. However in more onerous cases, with larger system sizes, we are likely going to require an automatic way to obtain $G,$ in a similar way to that shown for $s.$ Several promising approaches are documented in the literature for this \cite{Boccaletti1995, Lehnert:2011hu} and we are hopeful one will be beneficial for the fluid problem.

TDF has primarily been developed as a means to stabilise periodic orbits. While we have concentrated on steady and travelling wave solutions here, it is an obvious next step to consider UPOs. Preliminary results in this direction indicate that particular attention needs to be given to obtaining a highly accurate period in order for the method to be successful, as well as taking care of underlying symmetries and initial condition. {We hope to report progress on this in due course.} 

One very appealing feature of this work is the simplicity of the method and the ease with which it may be implemented. Any DNS code can be quite easily adapted to include the feedback term; the memory overhead associated with storing the history is not significant and is only slightly more onerous than required for the recurrent flow analysis. The nature of the method also makes it attractive as a way to `target' particular types of solution, particularly orbits which may be missed by recurrent flow analysis. This may lead to improved periodic orbit theory predictions when using the UPOs as a basis to recreate turbulent statistics \citep{Chandler, Cvitanovic:2013fz}. 

A direct comparison between TDF and recurrent flow analysis is not currently justifiable. While NGh clearly remains the best choice for computing UPOs, TDF has been shown, not only to effectively find travelling waves and equilibria, but importantly do so within a very simple framework. There is wide scope for TDF to become a powerful tool for studying nonlinear dynamics in fluid mechanics particularly given the extensive literature of extensions which may improve the method's efficiency further \citep{Schuster1997,Pyragas:2006gv,Flunkert:2011cg,Lehnert:2011hu,Pyragas:2014ba}.

\vspace{1cm}

\noindent
{\em Acknowledgements}. 

This work is supported by EPSRC New Investigator Award EP/S037055/1 ``Stabilisation of exact coherent structures in fluid turbulence''. We thank anonymous reviewers, Prof R. Kerswell \& Prof J. J. Healey for helpful comments on the manuscript.

\vspace{1cm}

 The authors report no conflict of interest.
 

\bibliography{papers}
\end{document}